\documentclass[aps,prd,longbibliography,nofootinbib]{revtex4-2}

\usepackage{graphicx}
\usepackage{amsmath,amssymb,bm}
\usepackage{siunitx}
\usepackage{booktabs}
\usepackage{microtype}
\usepackage{natbib}
\usepackage{hyperref}
\hypersetup{colorlinks=true,allcolors=blue}

\usepackage{array}
\usepackage{tabularx}
\usepackage{float}
\usepackage[section]{placeins}

\newcommand{\AU}{\mathrm{AU}}
\newcommand{\pcunit}{\mathrm{pc}}
\newcommand{\Rsun}{R_{\odot}}
\newcommand{\yr}{\mathrm{yr}}
\newcommand{\kms}{\mathrm{km\,s^{-1}}}
\newcommand{\AUyr}{\mathrm{AU\,yr^{-1}}}

\newcommand{\kWe}{\mathrm{kW_e}}
\newcommand{\MWe}{\mathrm{MW_e}}

\newcommand{\kWth}{\mathrm{kW_{th}}}
\newcommand{\MWth}{\mathrm{MW_{th}}}
\newcommand{\kgperkWe}{\mathrm{kg\,kW_e^{-1}}}
\newcommand{\gmsq}{\mathrm{g\,m^{-2}}}

\newcommand{\vinf}{v_{\infty}}
\newcommand{\vrbar}{\bar v_{r}}
\newcommand{\tbal}{t_{\rm bal}}
\newcommand{\trep}{t_{\rm rep}}
\newcommand{\tmission}{t_{\rm mission}}
\newcommand{\tinj}{t_{\rm inj}}
\newcommand{\tmargin}{t_{\rm margin}}
\newcommand{\tarr}{t_{\rm arr}}

\newcommand{\rp}{r_{\rm p}}
\newcommand{\sigmatot}{\sigma_{\rm tot}}
\newcommand{\chis}{\chi_{\rm s}}

\newcommand{\Pe}{P_{\rm e}}
\newcommand{\alphatot}{\alpha_{\rm tot}}
\newcommand{\alphaeff}{\alpha_{\rm eff}}
\newcommand{\Isp}{I_{\rm sp}}
\newcommand{\ve}{v_{\rm e}}

\newcommand{\Thrust}{F_{\rm EP}}

\newcommand{\mprop}{m_{\rm prop}}
\newcommand{\mpay}{m_{\rm pay}}
\newcommand{\mf}{m_{\rm f}}

\newcommand{\Tsail}{T_{\rm sail}}

\newcommand{\ie}{i.e.\ }

\newcommand{\Nrad}{N_{\rm rad}}           % number of radiating membrane/radiator sides
\newcommand{\etapc}{\eta_{\rm pc}}        % reactor-to-electric conversion efficiency
\newcommand{\gammarad}{\gamma_{\rm rad}}  % radiator areal mass density

\begin{document}

\title{Propulsion Trades for a 2035--2040 Solar Gravitational Lens Mission}

\author{Slava G. Turyshev}
\affiliation{
Jet Propulsion Laboratory, California Institute of Technology,\\
4800 Oak Grove Drive, Pasadena, CA 91109-0899, USA
}%

\date{\today}

\begin{abstract}
The Solar Gravitational Lens (SGL) enables resolved imaging and spectroscopy of nearby terrestrial exoplanets, but useful science begins only after a spacecraft reaches roughly \(650\)--\(900\) astronomical units (AU). A \(20~\yr\) lower-bound trip to \(650~\AU\) requires an average radial speed of \(32.5~\AU\) per year, or \(154~\kms\), before launch, targeting, steering, and operations margins. We compare close-perihelion solar sailing, fission-electric nuclear electric propulsion (NEP), and high-thrust Oberth injection followed by NEP cruise using common lower-bound outbound-leg architecture envelopes, not closed end-to-end trajectories. For an ideal sail passing \(0.05~\AU\) from the Sun, total sailcraft areal density must be about \(4.9\) grams per square meter to reach \(105~\kms\), and \(2.3\) grams per square meter to reach \(155~\kms\). Thus sub-\(20~\yr\) sail-only access requires ultra-low areal density plus deep-perihelion thermal qualification. For a \(20~\mathrm{t}\) NEP spacecraft with \(800~\mathrm{kg}\) payload and \(\Isp=9000~\mathrm{s}\), optimized constant-power transfers reach \(650~\AU\) in \(\trep\simeq27\)--\(33~\yr\) when the integrated power-plus-propulsion specific mass is \(10\)--\(20\) kg per electric kilowatt, requiring \(0.18\)--\(0.30\) megawatt-electric (\(\mathrm{MW_e}\)) and few-newton thrust. NEP-only \(\trep\le20~\yr\) requires \(\lesssim3\) kg per electric kilowatt, while hybrid architectures can approach \(\trep\simeq20~\yr\) if an upstream injection stage supplies \(50\)--\(70~\kms\). Thus sail-first is the nearer-term lightweight-access path; hybrid injection+\;NEP is higher-capability but requires prior high-energy-injection and \(0.2\)--\(0.4~\mathrm{MW_e}\) integrated NEP demonstrations.

\end{abstract}

\maketitle

%\tableofcontents

\section{Introduction}
\label{sec:intro}

The Solar Gravitational Lens (SGL), formed by the Sun's gravitational field, provides extreme gain and angular resolution that can enable direct multipixel imaging and spectroscopy of a nearby terrestrial exoplanet with meter-class telescopes \cite{Turyshev2026exoE,TuryshevMNRAS2022,TuryshevStarshade2025}.  Realistic mission concepts \cite{Turyshev-2020-NIAC_Phase2,Helvajian2022,Friedman_2024ExA} place science operations along the SGL focal line at heliocentric distances \(z\simeq 650\)--\(900~\AU\), where the spacecraft must (i) maintain precise pointing near the solar limb, (ii) execute controlled lateral motion in the image plane for pixel-by-pixel sampling and image reconstruction, and (iii) sustain power and communications where solar flux is \(\sim 4.2\times 10^{5}\) to \(8.1\times 10^{5}\) times weaker than at 1~AU. At these distances, solar power is not an enabling resource for high-rate communications or precision control; radioisotope or fission power is required throughout the science phase. Table~\ref{tab:notation} summarizes the notation, abbreviations, and electrical-power unit conventions used throughout the manuscript.

\begin{table*}[t]
\caption{Principal notation, abbreviations, and unit conventions used in the manuscript. Electrical and thermal powers are distinguished by subscripts \(e\) and \(th\), respectively; \(\mathrm{kW_e}\) and \(\mathrm{MW_e}\) denote electrical kilowatts and megawatts.}
\label{tab:notation}
\centering
\small
\begin{tabular}{lll}
\hline\hline
Symbol or unit & Meaning & Units or convention \\
\hline
\(z\) & Heliocentric distance along the SGL focal line & \(\AU\) \\
\(b\) & Solar impact parameter of the focused ray & m, \(\Rsun\) \\
\(\vinf\) & Heliocentric hyperbolic excess speed after inner-solar-system departure & \(\kms\), \(\AUyr\) \\
\(\vrbar\) & Mean radial speed required to reach \(z\) in a specified time & \(\kms\), \(\AUyr\) \\
\(\tbal\) & Ideal ballistic cruise time, \(z/\vinf\) & \(\yr\) \\
\(\trep\) & Reported outbound-leg lower-bound time used in this paper & \(\yr\) \\
\(\tmission\) & End-to-end mission-elapsed time & \(\yr\) \\
\(\tinj\) & Launch, targeting, solar-approach, and injection overhead & \(\yr\) \\
\(\delta_{\rm steer}\) & Effective fractional steering-loss penalty & dimensionless \\
\(\rp\) & Solar perihelion distance & \(\AU\), \(\Rsun\) \\
\(\sigmatot\) & Total sailcraft areal density, including sail, structure, power, payload, and bus & \(\gmsq\) \\
\(\chis\) & Effective sail pressure factor relative to an ideal radial perfect reflector & dimensionless \\
\(\Pe\) & Electrical power delivered to the electric-propulsion system & \(\kWe\), \(\MWe\) \\
\(\alphatot\) & Integrated stage-level specific mass of power plus propulsion system & \(\kgperkWe\) \\
\(\Isp\) & Specific impulse & s \\
\(\Thrust\) & Electric-propulsion thrust & N \\
\(\mprop\) & Propellant mass & kg, t \\
\(\pcunit\) & Parsec, \(1~\pcunit=3.086\times10^{16}~\mathrm{m}\) & distance unit \\
\(\kWe,\MWe\) & Kilowatt or megawatt of electrical power & electrical power \\
\(\kWth,\MWth\) & Kilowatt or megawatt of thermal power & thermal power \\
\hline\hline
\end{tabular}
\end{table*}

The SGL is scientifically compelling because it enables an observational regime that is otherwise far beyond credible near-term telescope or interferometer scaling: direct high-resolution imaging and spatially resolved spectroscopy of an Earth-like exoplanet at tens of parsecs. For non-SGL approaches as currently conceived, even a $\sim 10\times 10$ surface map of an exo-Earth at 10~pc is not achievable on human timescales, with representative mapping times reaching $\sim 10^{4}$--$10^{5}$ years for a 10~m class telescope once realistic backgrounds and systematics are included \cite{Turyshev2026exoE}. By contrast, the SGL provides enormous on-axis amplification ($\sim 10^{11}$) and extreme angular resolution ($\sim 10^{-10}$~arcsec), collapsing per-pixel integration times from centuries to hours and making $10^{4}$-pixel-class reconstructions feasible on mission-relevant timescales, enabling surface/atmosphere context, time variability (rotation, weather, seasons), and spatially localized interpretation of biosignature-relevant spectroscopy \cite{Turyshev2026exoE,TuryshevMNRAS2022,TuryshevStarshade2025}. This motivates treating transportation to $z\gtrsim 650$--$900~\AU$ not as ``deep space for its own sake,'' but as access to a unique astrophysical instrument.

For a 2035--2040 mission start, the dominant feasibility discriminator is outbound schedule to first science.
The basic kinematic requirement is severe: reaching \(z=650~\AU\) in \(20~\yr\) implies \(\vrbar\simeq 154~\kms\) even for an idealized ballistic cruise, and powered trajectories require still higher terminal speeds because a substantial fraction of the transfer occurs during acceleration. This schedule pressure makes propulsion an integrated system-closure problem rather than a simple \(\Delta v\) budget: thrust-to-mass, propellant throughput, lifetime, heat rejection, and flight-approval programmatics all affect feasibility. 

Historically, SGL mission architecture studies emphasized close-perihelion solar sailing because it can produce high \(\vinf\) with minimal propellant \cite{Helvajian2022}.
In parallel, NASA and DOE have renewed investments in space fission power and high-power electric propulsion for transport-class missions, motivating a fresh engineering assessment of whether nuclear electric propulsion (NEP) \cite{MasonTM2022,FSP}  can (i) reduce SGL transportation time, (ii) increase deliverable payload mass and electrical power at the focal region, or (iii) serve as the operations-enabling power+propulsion backbone even if it is not the fastest transportation mode.

The contribution of this paper is not a new propulsion principle or a high-fidelity optimal-control trajectory solver. Its contribution is a common lower-bound architecture-envelope method for SGL transportation. The method maps each propulsion family into the same mission discriminator---time to reach \(z=650\)--\(900~\AU\)---while retaining the hardware parameters that dominate feasibility: \((\sigmatot,\rp,\chis)\) for close-perihelion sailing, \((\alphatot,\Pe,\Isp,\mprop)\) for NEP, and injected speed \(v_0\) plus Oberth-stage mass ratio for hybrid architectures. This converts broad propulsion concepts into quantitative requirements on sail areal density, perihelion distance, stage specific mass, thrust-to-mass ratio, burn duration, total impulse, radiator area, and injection-stage burden.

The paper therefore makes three engineering contributions. First, it maps the SGL operating range \(z=650\)--\(900~\AU\) into required \(\vinf\), \(\vrbar\), and lower-bound transportation time. Second, it compares solar sailing, NEP, and Oberth-assisted hybrid injection using a common time-to-distance metric. Third, it exposes the system-closure quantities that determine whether each lower-bound architecture can plausibly mature by a 2035--2040 mission start: sail thermal/areal-density qualification, NEP stage specific mass, EP lifetime and throughput, heat rejection, and high-energy injection feasibility.

The analysis is intentionally first-order and comparative. Analytic and semi-analytic scalings are used to expose dominant dependencies and avoid hidden assumptions from black-box trajectory solvers. Accordingly, the quoted times are optimistic lower bounds: realistic mission designs must add architecture-dependent injection time, steering losses, and operational margin for long-duration propulsion and power.

This paper is organized as follows.
Section~\ref{sec:context} translates SGL operating distance (\(650\)--\(900~\AU\)) into kinematic requirements on \(\vinf\), \(\vrbar\), and lower-bound transportation time.
Section~\ref{sec:payload} summarizes payload-side drivers (telescope size, solar-light suppression, and the starshade option) that couple directly to delivered mass and power at the focal region.
Section~\ref{sec:model} presents the propulsion models used for the trade, including the sail \((\sigmatot,\rp)\rightarrow \vinf\) scaling and the constant-power NEP stage closure in terms of \(\alphatot\), \(\Isp\), and propellant fraction.
Section~\ref{sec:sail} evaluates close-perihelion solar sailing, emphasizing areal-density and thermal-survivability realism at the sail sizes implied by 25--35~yr and sub-20~yr access.
Section~\ref{sec:NEP} evaluates NEP-only transfers and hybrid injection+\;NEP architectures, and introduces system-closure checks (thrust-to-mass, total impulse, radiator area, and lifetime) needed for engineering credibility.
Section~\ref{sec:NTP} assesses NTP primarily as an Oberth-capable injection enabler rather than as a standalone SGL solution.
Section~\ref{sec:TRL} synthesizes technology readiness, development gates, and programmatics for a 2035--2040 start.
Section~\ref{sec:concl} concludes with an architecture recommendation tied explicitly to achievable TRL and schedule risk.

\section{Mission context and quantitative requirements}
\label{sec:context}

\subsection{SGL focal region and practical operating range}

In the solar gravitational monopole approximation, the start of the SGL focal line, $z_0$, for a ray with impact parameter $b$ occurs at \cite{Turyshev-2017,Turyshev-2019}
\begin{equation}
z_0 \simeq \frac{b^2}{2 r_g}, \qquad {\rm where} \qquad r_g \equiv \frac{2GM_\odot}{c^2}\simeq \SI{2.95}{km}.
\end{equation}
For $b \approx \Rsun$, $z_0 \approx \SI{548}{\AU}$, consistent with detailed SGL analyses and mission studies \cite{Helvajian2022}. Practical observing with meter-class apertures and internal coronagraphs typically adopts
$z \gtrsim \SI{650}{\AU}$ to increase angular separation between the solar disk and the Einstein ring
\cite{TuryshevMNRAS2022,TuryshevStarshade2025}.
Architectures using an external occulter and/or different stray-light assumptions may begin useful
science closer to the start of the focal line (e.g., $z\sim 550$--$650~\AU$), but we retain 650--900 AU as a conservative, coronagraph-compatible baseline for this trade.

\subsection{Image-plane scale drives navigation and power, not propulsion}

For an exo-Earth observed through the SGL, the projected image in the SGL image plane spans kilometer scales for targets at $\sim 10$--$30$~pc \cite{TuryshevMNRAS2022,TuryshevStarshade2025}.
A representative mission-architecture study reports pixel-step scales of order $\sim \SI{10}{m}$ and lateral navigation accuracy requirements at the $\sim \SI{0.1}{m}$ level for robust image reconstruction \cite{Helvajian2022}.
These requirements drive onboard power, autonomy, and propulsion for transverse control during operations, but do not materially change the outbound $\Delta v$ requirement, which is dominated by the heliocentric cruise distance.

\subsection{Outbound speed required for $\le 30$~yr and $\le 20$~yr access}
\label{sec:TOF-20-30}

\begin{figure}[t]
\includegraphics[width=0.65\linewidth]{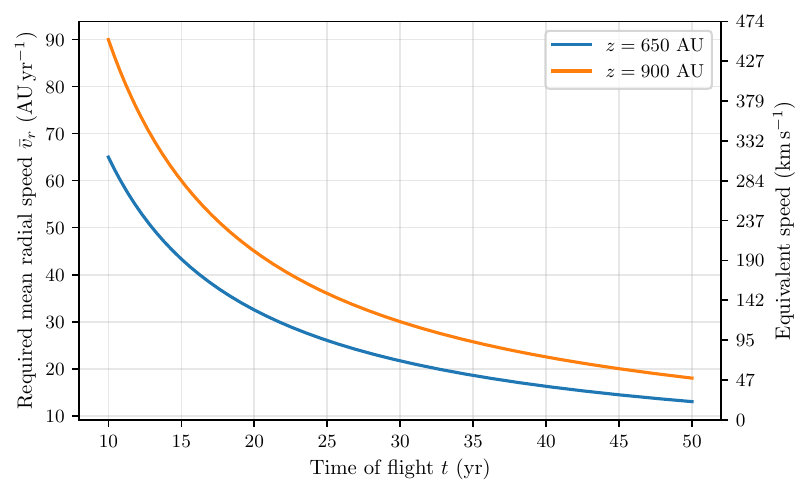}
\caption{Required mean radial speed versus travel time for reaching \(z=650\) and \(900~\AU\). The \(20~\yr\) lower-bound requirement corresponds to \(\bar v_r\simeq32.5~\AUyr\simeq154~\kms\) to
\(650~\AU\), before architecture-dependent injection and steering
penalties are included.}
\label{fig:reqspeed}
\end{figure}

A useful first-order discriminator among propulsion options is the heliocentric hyperbolic excess speed $v_\infty$ after leaving the inner solar system. Neglecting gravity losses and assuming a near-radial outbound leg, the ideal ballistic cruise time is
\begin{equation}
\tbal(z,\vinf) \simeq \frac{z}{\vinf}.
\label{eq:tbal}
\end{equation}
with $z$ in $\AU$ and $v_\infty$ in $\AUyr$.
Table~\ref{tab:ballistic} lists representative values; Fig.~\ref{fig:reqspeed} shows the corresponding required mean radial speed for $z=650$ and $900~\AU$.

\begin{table}[t]
\caption{Ballistic cruise time to the SGL operating region as a function of asymptotic heliocentric speed $v_\infty$. Times neglect gravity losses and assume a nearly radial outbound trajectory after departure.}
\label{tab:ballistic}
%\begin{ruledtabular}
\begin{tabular}{cccc}
\hline
$v_\infty$ ($\AUyr$) & $v_\infty$ ($\kms$) & \(\tbal\) to \(650~\AU\) (\(\yr\)) & \(\tbal\) to \(900~\AU\) (\(\yr\))\\
\hline\hline
10 & 47.4 & 65.0 & 90.0\\
12 & 56.9 & 54.2 & 75.0\\
15 & 71.1 & 43.3 & 60.0\\
20 & 94.8 & 32.5 & 45.0\\
25 & 118.5 & 26.0 & 36.0\\
\hline
\end{tabular}
%\end{ruledtabular}
\end{table}

For example, a purely ballistic trajectory that cruises at
\(20~\AUyr\) requires \(650/20\simeq32.5~\yr\) to reach
\(650~\AU\). Meeting a $20$-year requirement to $650~\AU$ demands $\sim 32.5~\AUyr$ mean radial speed, and powered trajectories take longer than $650~\AU/\vinf$ because a significant fraction of the distance is covered while the vehicle is still accelerating.

A mission seeking first science return in \(\lesssim30~\yr\) must
therefore target \(\vinf\gtrsim20~\AUyr\simeq95~\kms\). A mission seeking arrival in $\lesssim \SI{20}{\yr}$ requires $\bar v_r \simeq 32.5~\AUyr \simeq \SI{154}{\kms}$ even under idealized ballistic assumptions, and in practice requires still higher speed to accommodate inner-solar-system injection time.

\subsection{Reported outbound-leg time versus mission-elapsed time}
\label{sec:tof_scope}

Unless stated otherwise, all transfer times in this paper are reported outbound-leg lower-bound times, denoted \(\trep\). The clock for \(\trep\) starts after the architecture-dependent launch, targeting, solar-approach, escape, and injection sequence has established the initial condition of the modeled long-range heliocentric leg. Thus, for close-perihelion sail and Oberth-enabled cases, \(\trep\) excludes the time to target perihelion and execute perihelion operations. For NEP-only cases, \(\trep\) excludes launch escape and the inner-solar-system spiral or targeting phase needed to establish the modeled outbound state.

The end-to-end mission-elapsed time is therefore
\begin{equation}
\tmission =
\tinj + (1+\delta_{\rm steer})\trep + \tmargin ,
\label{eq:tmission}
\end{equation}
where \(\tinj\) collects architecture-dependent launch, targeting, solar-approach, escape, and injection overhead, \(\delta_{\rm steer}\) is an effective fractional steering-loss penalty, and \(\tmargin\) is operational margin for long-duration propulsion, power, navigation, and fault recovery.

For low-thrust NEP, \(v_0\simeq0\) in the lower-bound model should not be interpreted as radial thrusting from rest at \(1~\AU\). It denotes the initial speed of the idealized rectilinear outbound model after launch/escape and any inner-solar-system targeting have been absorbed into \(\tinj\). The reference acceleration scale is
\begin{equation}
a_0 =
\frac{2\eta\Pe}{m_0g_0\Isp}
\simeq
2.4\times10^{-4}~\mathrm{m\,s^{-2}}
\left(\frac{\Pe}{300~\kWe}\right)
\left(\frac{20~\mathrm{t}}{m_0}\right)
\left(\frac{9000~\mathrm{s}}{\Isp}\right).
\label{eq:nep_a0_context}
\end{equation}
The solar gravitational acceleration equals this value at
\begin{equation}
r_\times \simeq
1~\AU
\left[
\frac{g_\odot(1~\AU)}{a_0}
\right]^{1/2}
\simeq
5.0~\AU
\left(\frac{2.4\times10^{-4}~\mathrm{m\,s^{-2}}}{a_0}\right)^{1/2}.
\label{eq:nep_gravity_crossover}
\end{equation}
Cases that numerically achieve \(\trep\simeq20~\yr\) should therefore be interpreted as lower-bound requirement boundaries, not as complete \(20~\yr\) mission designs.

\subsection{Why chemical propulsion is insufficient for SGL access}
\label{sec:chem_insufficient}

The speeds required for $\lesssim 20$--$30~\yr$ access to $z\gtrsim 650~\AU$
($v_\infty \sim 95$--$155~\kms$) are far beyond what chemical propulsion can deliver
at useful payload mass.
With state-of-the-art chemical specific impulse $I_{sp}\lesssim 450~\mathrm{s}$
($v_e=g_0 I_{sp}\lesssim 4.4~\kms$), the ideal rocket equation implies mass ratios
$m_0/m_f=\exp(\Delta v/v_e)\gtrsim 10^{9}$--$10^{15}$ for $\Delta v=95$--$155~\kms$,
before accounting for inert mass and margins \cite{SuttonBiblarzRPE9,BateMuellerWhite1971}.
Even when gravity assists are exploited, the fastest deep-space missions achieve only
\(\sim 3\)--\(4~\AUyr\)-class heliocentric escape speeds, i.e., an order of magnitude below the
SGL requirement \cite{NASA_VoyagerFactSheet2025,JPL_NewHorizonsOverview2026}.
Deep-solar Oberth maneuvers reduce the required impulsive $\Delta v$, but the remaining
chemical requirement is still prohibitive at realistic stage mass fractions: at $r_p=0.02~\AU$,
achieving $v_\infty\simeq 95~\kms$ requires $\Delta v\simeq 15~\kms$ (ideal $m_0/m_f\simeq 30$),
while achieving $v_\infty\simeq 155~\kms$ requires $\Delta v\simeq 38~\kms$, which is infeasible
for chemical stages.

Finally, flight heritage provides a reality check: even with gravity assists, the fastest chemical/gravity-assist deep-space missions have departed the solar system at only \(\sim3\)--\(4~\AUyr\)
(\(\sim15\)--\(20~\kms\)) heliocentric hyperbolic speed \cite{NASA_VoyagerFactSheet2025,JPL_NewHorizonsOverview2026}, an order of magnitude below the \(\gtrsim20\)--\(30~\AUyr\) class speeds implied by SGL timelines.

\section{Payload implications: telescope size and sunlight suppression}
\label{sec:payload}

\subsection{Internal coronagraph constraint and wavelength reach}

For an internal coronagraph, the telescope diffraction limit constrains the ability to resolve the Einstein ring from the solar disk.
Detailed SGL sensitivity calculations show that for a $d=\SI{1}{m}$ telescope, the beginning of science operations is pushed to $z \gtrsim \SI{650}{\AU}$ for $\lambda\sim \SI{1}\mu{\rm m}$, while mid-IR ($\lambda\sim \SI{10}\mu{\rm m}$) operation would demand an impractically large aperture ($d\sim \SI{15}{m}$) \cite{TuryshevStarshade2025}.

\subsection{External occulter (starshade) decouples wavelength reach from aperture}

An external occulter placed at separation $z_s$ from the telescope and of diameter $D_0$ blocks the solar disk without being limited by telescope diffraction. A representative configuration is a $\sim \SI{70}{m}$ starshade at $z_s\sim \SI{5000}{km}$, which can block the Sun with Fresnel number $F\sim 25$ at $\lambda=\SI{10}\mu{\rm m}$ \cite{TuryshevStarshade2025}. With a starshade, a \(d=0.4~\mathrm{m}\) telescope can achieve broadband signal-to-noise ratio (SNR) sufficient for image reconstruction across \(0.1\)--\(20~\mu\mathrm{m}\) for an exo-Earth at \(30~\pcunit\) observed from \(z=650~\AU\) \cite{TuryshevStarshade2025}. This relaxes the requirement to deliver large apertures to \(650\)--\(900~\AU\) and is therefore directly relevant to the propulsion trade. The starshade linear dimension is of the same order as some proposed SGL sailcraft structures, but the optical-figure, edge-scatter, deployment, and formation-flying requirements are different from those of a propulsion sail. Any sail-as-occulter reuse must therefore be treated as a separate mass, optical, thermal, and control closure rather than as a free secondary function.

\section{Propulsion models used for engineering trades}
\label{sec:model}

We use lightweight engineering models designed to (i) preserve the dominant scaling of \(\trep\)  with key design variables and (ii) expose requirements on mass, power, and thermal management. They are not full optimal-control trajectory solutions; gravity and steering losses are neglected in the long-distance outbound leg, which is appropriate for comparative trades at the $\sim 10$--$40$~yr level.

\subsection{Interpretation of lower-bound model outputs}
\label{sec:lower_bound_assumptions}

The models below are intentionally lower-bound models. They are useful for comparing propulsion-family access envelopes, but they do not by themselves close a mission design. Table~\ref{tab:lower_bound_assumptions} summarizes the dominant idealizations, their bias direction, and the quantitative interpretation used in the paper.

\begin{table*}[t]
\caption{Dominant lower-bound assumptions and their quantitative interpretation. These effects are not hidden in the architecture comparison; they define the gap between the reported outbound-leg time \(\trep\) and the end-to-end mission time \(\tmission\).}
\label{tab:lower_bound_assumptions}
\centering
\small
\begin{tabular}{lll}
\hline\hline
Assumption & Optimistic bias & Quantitative interpretation in this paper \\
\hline
Reported times exclude launch, tar- & \(\trep\) is shorter than mission-elapsed time & End-to-end time is represented by Eq.~(\ref{eq:tmission}); \\
geting, solar approach, and injec-& & \(\tinj\) must be supplied by an architecture- \\
tion overhead & & specific launch and injection design. \\

Near-radial long-distance outbound  & Neglects steering losses, gravity losses,   & Steering and non-radial thrusting enter as  \\

leg &  and angular-momentum removal &  \((1+\delta_{\rm steer})\trep\). A \(10\%\) steering penalty in-\\
 &   & creases a nominal \(20~\yr\) lower bound by \(2~\yr\). \\

Ideal sail optical response, \(\chis=1\) & Overestimates sail acceleration and \(\vinf\) & At fixed \((\sigmatot,\rp)\), \(\vinf\propto\chis^{1/2}\). Thus  \\
&  &  \(\chis=0.8\) reduces speed by \(10.6\%\) and  \\
&  &  increases ballistic time by about \(12\%\). \\

Constant-power NEP stage with  & Omits launch escape, spiral-out, throttling,& NEP-only \(v_0\simeq0\) cases are post-escape lower  \\

 1D dynamics &  solar gravity during escape, and trajectory  & bounds, not radial thrust from rest at \(1~\AU\). \\
   &   steering &  \\

Integrated NEP specific mass & Fixed masses, margins, tankage, and feed  & The effective value \(\alphaeff\) should include  \\

represented by \(\alphatot\) &  systems can be underrepresented & growth and fixed dry masses; Eq.~(\ref{eq:alpha_tank_penalty}) shows  \\
 &  &  that \(5\%\) tankage on \(16~\mathrm{t}\) propellant adds \ \\
  &  &   \(\simeq2.7~\kgperkWe\) at \(300~\kWe\). \\

Ideal radiator and thermal   & Underestimates radiator packaging, degra-& Radiator mass consumes several \(\kgperkWe\)  \\

rejection estimates & dation, shielding, and view-factor penalties & of the \(\alphatot=10\)--\(20~\kgperkWe\) budget  \\
 
   && even before reactor, PMAD, PPU, and EP  \\
   && redundancy. \\

Impulsive solar-Oberth injection & Underestimates finite-burn, thermal-protec-& Oberth tables report ideal mass ratios only;  \\

& tion, cryogenic, targeting, and perihelion- &  they are requirement boundaries on the  \\

&operation penalties &  injection architecture, not closed injection-\\

& & stage designs. \\

\hline\hline
\end{tabular}
\end{table*}

\subsection{Solar sail photonic-assist scaling}

For a sailcraft leaving the inner solar system after a deep solar perihelion pass, a useful upper-bound scaling for the post-perihelion hyperbolic speed is \cite{Turyshev-2020-NIAC_Phase2,Friedman_2024ExA,Helvajian2022,Sundivers_2023}
\begin{equation}
\vinf \simeq
\left(\frac{2\mu_\odot \beta}{\rp}\right)^{1/2},
\qquad
\beta \equiv \frac{a_0}{g_\odot(1~\AU)} .
\label{eq:sail_vinf}
\end{equation}
For a perfectly reflecting sail at 1~AU, \(p_0=2S_0/c\simeq 9.08~\mu\mathrm{N\,m^{-2}}\). To retain nonideal optical properties and non-radial steering in a single first-order parameter, define \(\chis\equiv p_{\rm eff}/p_0\), with \(\chis=1\) for the ideal radial, perfectly reflecting case. Then
\begin{equation}
a_0=\frac{\chis p_0}{\sigmatot},
\qquad
\beta =
\frac{\chis p_0}{\sigmatot g_\odot(1~\AU)}.
\label{eq:sail_beta_nonideal}
\end{equation}
Equivalently, the required total system areal density is
\begin{equation}
\sigmatot \simeq
4.93~\chis
\left(\frac{105~\kms}{\vinf}\right)^2
\left(\frac{0.05~\AU}{\rp}\right)
~\mathrm{g\,m^{-2}} .
\label{eq:sail_sigma_req}
\end{equation}
Thus the Fig.~\ref{fig:sailsigma} curves are ideal-\(\chis=1\) boundaries: if \(\chis=0.8\), the same physical sail reaches only \(0.8^{1/2}=0.894\) of the ideal \(\vinf\), or must reduce \(\sigmatot\) by 20\% to recover the same \(\vinf\).

For comparison, maintaining \(\sigmatot \simeq 5~\mathrm{g\,m^{-2}}\) with a \(\sim 10^3\)-kg payload implies sail areas of order \(10^5~\mathrm{m^2}\) (hundreds of meters on a side). Fast sail-first architectures therefore favor smaller spacecraft, distributed systems, or both, and are not payload-equivalent to the \(m_0=20~\mathrm{t}\), \(\mpay=800~\mathrm{kg}\) NEP reference case.

A solar sail can provide propellantless acceleration in the inner solar system, but it does not provide electrical power.
A sailcraft must therefore carry a non-solar power source---most conservatively a radioisotope power system
(RPS; \ie an RTG-class unit for $\mathcal{O}(10^2~\mathrm{W_e})$) or a compact fission system---to support avionics, attitude control, autonomy, and telecommunications during the multi-decade cruise and at the SGL focal region. Because sail performance scales with the \emph{total} system areal density $\sigmatot$, the mass of the power system must be included in $\sigmatot$ and can materially change the required sail size or achievable $v_\infty$ (Secs.~\ref{sec:model} and \ref{sec:sail}).

\subsection{Constant-power vacuum NEP model}
\label{sec:nep_model}

The NEP equations used here are not launch-vehicle or atmospheric-flight equations. They model an in-space electric-propulsion stage after launch and escape have established the initial state of the modeled outbound leg. Their purpose is to connect electrical power, specific impulse, propellant throughput, and stage mass to the time required to reach the SGL operating region.

For an electric thruster operating at electrical power \(\Pe\), efficiency
\(\eta\), and exhaust velocity \(\ve=g_0\Isp\), the ideal power balance is
\begin{equation}
\eta\Pe = \frac{1}{2}\dot m \ve^2 .
\label{eq:nep_power_balance}
\end{equation}
The corresponding vacuum thrust and propellant mass flow are
\begin{equation}
\Thrust = \dot m \ve
= \frac{2\eta\Pe}{g_0\Isp},
\qquad
\dot m =
\frac{\Thrust}{g_0\Isp}
=
\frac{2\eta\Pe}{g_0^2\Isp^2}.
\label{eq:nep_thrust_mdot}
\end{equation}
For constant \(\Pe\) and \(\Isp\), the ideal velocity increment is
\begin{equation}
\Delta v = g_0\Isp
\ln\!\left(\frac{m_0}{\mf}\right).
\label{eq:nep_rocket_equation}
\end{equation}

A central system variable is the integrated stage-level specific mass \(\alphatot\), which includes the reactor and shielding, power conversion, heat rejection, power management and distribution (PMAD), power processing unit (PPU), thrusters, gimbals, harnessing, and representative redundancy. The dry-mass closure used for the lower-bound trades is
\begin{equation}
m_{\rm dry}=\mpay+\alphatot\Pe,
\label{eq:dry_mass}
\end{equation}
with wet mass
\begin{equation}
m_0=m_{\rm dry}+\mprop .
\label{eq:wet_mass}
\end{equation}
Here \(\mpay\) includes the telescope, communications, structure, and non-propulsive spacecraft subsystems, while \(\mprop\) is the electric-propulsion propellant mass. Time-to-distance is evaluated using the analytic constant-thrust integration in Appendix~\ref{app:nep_time_model}.

In this paper, $\alphatot$ is intended as an \emph{integrated} specific mass for the flight stage and therefore
implicitly subsumes the mass of the power source (reactor and shield), conversion, heat rejection, PMAD, PPU,
and the propulsion string (thrusters, gimbals, harnessing, and representative redundancy). Propellant tankage and feed-system mass can be treated either as part of $\mpay$ (spacecraft bus) or absorbed into an effective $\alphatot$; because the optimized cases already require large propellant fractions (Tables~\ref{tab:nep_points}--\ref{tab:nep_sens}), including realistic tankage would modestly increase $\alphatot$ and therefore lengthen \(\trep\) relative to the optimistic lower bounds reported here.

Because $\alphatot$ is not strictly power-invariant, a useful first-order decomposition is
\begin{equation}
m_{\rm ps} \;=\; m_{\rm fix} + \alpha_{\rm var}\,\Pe,
\qquad
\alphatot \equiv \frac{m_{\rm ps}}{\Pe} \;=\; \alpha_{\rm var} + \frac{m_{\rm fix}}{\Pe},
\label{eq:alpha_scaling}
\end{equation}
where $m_{\rm fix}$ aggregates weakly-scaling items (structure, controls, margins, shielding geometry constraints)
and $\alpha_{\rm var}$ aggregates approximately power-proportional items (conversion hardware, radiator area/mass
at fixed $T_{\rm rad}$, PMAD, and EP string scaling).
Eq.~(\ref{eq:alpha_scaling}) makes explicit why $\alphatot$ values inferred from MW-class point designs do not automatically translate to $\Pe\sim 0.2$--$0.4~\mathrm{MW_e}$ stages, and why uncrewed SGL layouts (larger boom separation, shadow shields, relaxed dose limits) can materially improve $\alphatot$ compared to crewed transports.

\paragraph{Stage-level specific mass used here.}
Published nuclear-electric ``specific mass'' values are often quoted for the
fission-power subsystem only (reactor, shield, conversion, heat rejection, and
PMAD), and may also be reported at current-best-estimate without explicit
programmatic contingency. In contrast, the present trade variable \(\alphatot\)
is a flight-stage closure parameter that includes both the power subsystem and the
EP segment (PPU, thrusters, feed, harnessing, gimbals, and representative
redundancy). Accordingly, literature comparisons should be converted to an
integrated stage-level \(\alphatot\) before being mapped onto the contours used
here.

\paragraph{Effective comparison metric.}
To map subsystem current-best-estimate values into the present trades, we define
\begin{equation}
\alpha_{\rm eff}
=
\alpha_{\rm CBE}(1+\delta_{\rm mg})
+
\frac{m_{\rm tank}+m_{\rm feed}+m_{\rm margin}}{\Pe},
\label{eq:alpha_eff}
\end{equation}
where \(\delta_{\rm mg}\) is a mass-growth allowance and the second term collects fixed dry masses that do not scale directly with power. The contours reported in this paper should therefore be interpreted as optimistic lower-bound contours in
\(\alpha_{\rm eff}\)-space rather than as contingency-inclusive point designs.

A useful scale for interpreting Eq.~(\ref{eq:alpha_eff}) is the tankage/feed penalty associated with the large optimized propellant masses:
\begin{equation}
\Delta \alpha_{\rm tank}
\simeq
\frac{\epsilon_{\rm tank}\mprop}{\Pe}
=
2.7~\kgperkWe
\left(\frac{\epsilon_{\rm tank}}{0.05}\right)
\left(\frac{\mprop}{16~\mathrm{t}}\right)
\left(\frac{300~\kWe}{\Pe}\right).
\label{eq:alpha_tank_penalty}
\end{equation}
Thus even a 5\% inert allowance on a \(\sim 16~\mathrm{t}\) xenon system moves an apparent \(\alphatot=10~\kgperkWe\) design toward \(\alpha_{\rm eff}\simeq 13~\kgperkWe\), before adding trajectory steering losses or programmatic mass growth.

\paragraph{Comparison convention.}
The sail and NEP cases are not payload-equivalent point designs: representative
sailcraft are \(10^2\)-kg-class systems whose performance is set primarily by
\((\sigmatot,\rp)\), whereas the NEP reference case is a \(20\)-t spacecraft with
\(\mpay=800~\mathrm{kg}\) and long-lived electrical power. The present comparison should therefore be read as a propulsion-family access-envelope comparison rather than as an equal-capability observatory comparison.

\subsection{NTP impulsive and Oberth approximations}

NTP is modeled as an impulsive burn with $I_{sp}\sim 850$--$950$~s. If executed at a perihelion where the local speed is $v_p$, the approximate post-burn solar hyperbolic excess is
\begin{equation}
v_\infty \simeq \sqrt{2 v_p \Delta v + \Delta v^2}.
\label{eq:oberth}
\end{equation}
Eq.~(\ref{eq:oberth}) highlights why deep solar perihelia can multiply the effectiveness of high-thrust propulsion; the practical question is whether the thermal, cryogenic, and operations constraints of $r_p\lesssim 0.02~\AU$ are compatible with an NTP stage.

\section{Solar sailing to the SGL}
\label{sec:sail}

\subsection{Performance requirements: $\sigmatot$ versus $r_p$}

\begin{figure}[t]
\includegraphics[width=0.65\linewidth]{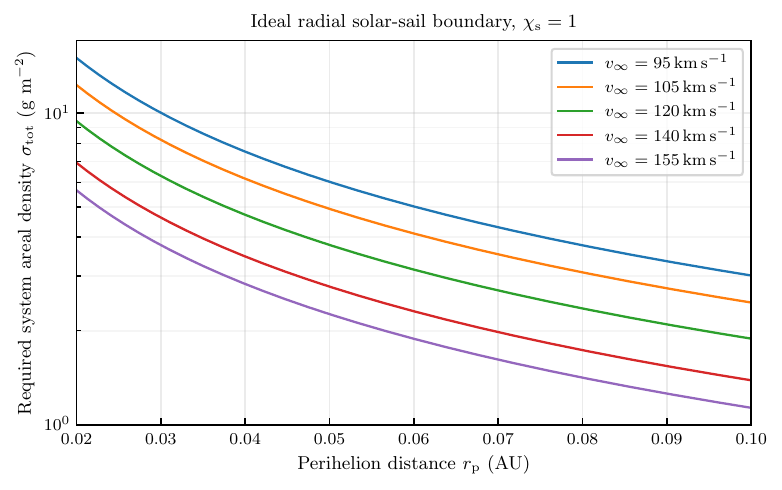}
\caption{Idealized system areal density \(\sigmatot\) required to achieve target solar hyperbolic excess speed \(\vinf\) using
Eqs.~(\ref{eq:sail_vinf})--(\ref{eq:sail_sigma_req}) with \(\chis=1\).
Nonideal optical response, coating degradation, or steering losses enter through \(\chis<1\); at fixed \((\sigmatot,\rp)\), the achieved \(\vinf\) scales as \(\chis^{1/2}\).}
\label{fig:sailsigma}
\end{figure}

Eq.~(\ref{eq:sail_vinf}) indicates that $v_\infty \sim \SI{100}{\kms}$ is accessible if $\beta$ is a few tenths and $r_p \sim 0.04$--$0.06~\AU$ ($\sim 8$--$13~\Rsun$).
Figure~\ref{fig:sailsigma} maps the required system areal density $\sigmatot$ versus perihelion for several target $v_\infty$ values. 

Table~\ref{tab:sail_area_scale} shows the corresponding sail-area scale at \(\rp=0.05~\AU\) for two schedule-relevant speeds. The comparison emphasizes that fast sail-first cases are not payload-equivalent to the \(m_0=20~\mathrm{t}\), \(\mpay=800~\mathrm{kg}\) NEP reference case.

\begin{table}[t]
\caption{Ideal sail area implied by Eq.~(\ref{eq:sail_sigma_req}) at \(\rp=0.05~\AU\) and \(\chis=1\). Square-side values are shown in parentheses.}
\label{tab:sail_area_scale}
\centering
\small
\begin{tabular}{cccc}
\hline\hline
Target class & \(\vinf\) & \(\sigmatot\) & \(A=m/\sigmatot\) \\
 & (\(\kms\)) & (\(\gmsq\)) & \(m=100~\mathrm{kg}\) / \(800~\mathrm{kg}\) \\
\hline
30~yr-class & 105 & 4.93 &
\(2.0\times 10^4~\mathrm{m^2}\) (143~m) /
\(1.6\times 10^5~\mathrm{m^2}\) (403~m) \\
20~yr lower bound & 155 & 2.26 &
\(4.4\times 10^4~\mathrm{m^2}\) (210~m) /
\(3.5\times 10^5~\mathrm{m^2}\) (595~m) \\
\hline\hline
\end{tabular}
\end{table}

Recent ``extreme solar sailing'' studies emphasize that very fast transits are achievable in principle only by combining ultra-low total areal density with very deep perihelia (a few solar radii), which moves the feasibility question from trajectory mechanics to coupled materials, thermal, and large-area deployment qualification. For example, \cite{DavoyanNIAC2021} analyzed extreme-proximity solar sailing ($\lesssim 5~\Rsun$) and discussed candidate metamaterial sail approaches together with the associated environmental and system challenges at these perihelia. These results reinforce the conclusion here: sub-20~yr sail-only access is not ruled out by physics, but it lives in a tightly coupled materials+structures+thermal qualification regime at mission scale.

\subsection{Non-solar electrical power is a first-order sail sizing driver}
\label{sec:sail_power}

Although the solar sail provides outbound propulsion, it does not provide electrical power at $z\gtrsim 650~\AU$ where solar flux is negligible. A sailcraft must therefore carry a non-solar power system---typically a radioisotope power system (RPS/RTG) or a small fission source---to support communications, attitude control, autonomy, and (for SGL operations) image-plane navigation and transverse propulsion.

This requirement fundamentally couples to transportation performance because the power-system mass enters the total system areal density:
\begin{equation}
\sigmatot =
\frac{m_{\rm sail}+m_{\rm struct}+\mpay+m_{\rm pwr}}{A_{\rm sail}},
\qquad
m_{\rm pwr}=\alpha_{\rm pwr} P_{\rm bus},
\label{eq:sail_sigma_total}
\end{equation}
or, equivalently,
\begin{equation}
\sigmatot = \sigma_0 + \sigma_{\rm pwr}
= \sigma_0 + \frac{\alpha_{\rm pwr}P_{\rm bus}}{A_{\rm sail}},
\label{eq:sail_sigma_pwr}
\end{equation}
where \(\sigma_0\) denotes the sailcraft areal density excluding the non-solar power subsystem. For illustration, a \(P_{\rm bus}=1~\kWe\) non-solar power system with \(\alpha_{\rm pwr}=100~\kgperkWe\) contributes \(m_{\rm pwr}=100~\mathrm{kg}\), which at \(\sigma_0=5~\mathrm{g\,m^{-2}}\) corresponds to an added area scale \(\Delta A\simeq 2\times 10^4~\mathrm{m^2}\).
Therefore, improvements in power specific mass (W\,kg$^{-1}$) or reductions in required electrical power relax the $(\sigmatot,r_p)\rightarrow \vinf$ requirements shown in Fig.~\ref{fig:sailsigma}, while higher power demand or heavier power systems push the design toward larger sails, deeper perihelia, or reduced payload.

Sail-first architectures still require continuous electrical power for guidance, navigation and control,
telecommunications, and onboard autonomy during a multi-decade cruise and at $z\gtrsim 650~\AU$, where solar arrays are infeasible. Representative SGL sailcraft concepts assume $\sim 100$--$150~\mathrm{W_e}$-class onboard power per vehicle. For reference, NASA's current Multi-Mission Radioisotope Thermoelectric Generator (MMRTG) \cite{NASA_SpaceNuclearSystems2022} produces $\sim 110~\mathrm{W_e}$ and has mass $\lesssim 45~\mathrm{kg}$. For $m\sim 100$--$200~\mathrm{kg}$ sailcraft, an RTG-class unit can therefore be a $\sim 20$--$50\%$ mass fraction and must be counted in $\sigmatot$. Since $v_\infty \propto \beta^{1/2} \propto \sigmatot^{-1/2}$ Eq.~(\ref{eq:sail_vinf}), neglecting RTG mass can overstate achievable $v_\infty$ by $\mathcal{O}(10$--$20\%)$ (or equivalently understate required sail area by tens of percent). Accordingly, all sail performance points in Fig.~\ref{fig:sailsigma} should be interpreted as requiring that $\sigmatot$ already includes the mass of the selected non-solar power system and associated spacecraft subsystems.

\subsection{Thermal environment at deep perihelion}

Deep perihelia are enabled only if the sail survives the solar heat flux. A thin-membrane equilibrium estimate is
\begin{equation}
\Tsail(r) \simeq
\left[
\frac{\alpha_{\rm abs}S_0(1~\AU/r)^2}
{N_{\rm rad}\epsilon\sigma_{\rm SB}}
\right]^{1/4}.
\label{eq:sail_temp}
\end{equation}
where \(S_0=1361~\mathrm{W\,m^{-2}}\), \(\alpha_{\rm abs}\) is absorptivity, \(\epsilon\) emissivity, and \(\Nrad=2\) for two-sided membrane radiation. For \(\alpha_{\rm abs}=0.05\), \(\epsilon=0.8\), and \(N_{\rm rad}=2\),
Eq.~(\ref{eq:sail_temp}) gives \(\Tsail\simeq740~\mathrm{K}\) at
\(r=0.05~\AU\) and \(\Tsail\simeq830~\mathrm{K}\) at \(r=0.04~\AU\). The temperature scales as \((\alpha_{\rm abs}/\epsilon)^{1/4}r^{-1/2}\), so modest coating degradation maps directly into perihelion or \(\sigmatot\) margin.

A mission-architecture study reports a representative (unoptimized) $\sim\SI{50}{kg}$ sailcraft with a $\sim10^{4}\,\mathrm{m^2}$-class sail achieving $21~\AUyr$ with $r_p=0.025~\AU$ at a maximum sail temperature of $\SI{950}{K}$ for reflectivity $\rho=0.95$ and emissivity $\epsilon=0.80$ \cite{Helvajian2022}.  This result is not directly scalable to heavier payloads without changing $\sigmatot$, but it anchors the thermal regime for ``fast'' sails.

\subsection{Solar sail realism for a 2035 start}

Solar sailing has direct flight heritage (IKAROS, LightSail-class missions) and active NASA development (ACS3) \cite{IKAROS,ACS3}. However, SGL-relevant performance depends on scaling from $\sim 10$~m-class demos to $\sim 150$--$300$~m sails while holding $\sigmatot$ to a few $\si{g\,m^{-2}}$, and on survival at $r_p \lesssim 0.1~\AU$. The mission-architecture study explicitly reports high TRLs (8--9) for several sailcraft subsystems but low TRL for the sail \emph{material system} needed for deep-perihelion operation \cite{Helvajian2022,Sundivers_2023}. This maturity profile is central to propulsion selection in the 2035--2040 window.

To avoid conflating distinct regimes, it is useful to separate \emph{(i) 25--40 yr-class sail access} from
\emph{(ii) sub-20 yr sail access}. The 25--40 yr regime corresponds to $\vinf\sim 70$--$110~\kms$ with $\sigmatot$ of a few~g~m$^{-2}$ and perihelia $\rp\sim 0.05$--$0.1~\AU$, where the primary challenges are large-area deployment dynamics, metrology, and maintaining optical properties at elevated temperature. In contrast, sub-20 yr sail-only access requires simultaneously deeper perihelia and substantially lower $\sigmatot$ (Fig.~\ref{fig:sailsigma}), pushing the design into a combined materials+structures+thermal corner where system-level maturity is presently lower. Thus, the ``low-maturity'' statement in this section should be interpreted as applying to the \emph{sub-20 yr} sail-only regime at mission scale, not to solar sailing as a whole.

A useful anchor on realistic readiness is provided by recent mission studies that identify areal density and perihelion survivability as the key discriminators for fast deep-space applications, together with near-term flight demonstrations such as ACS3. These studies consistently emphasize that sub-$20~\yr$ sail-only access requires pushing the combined $(\sigmatot,r_p)$ design space (Fig.~\ref{fig:sailsigma}), i.e., achieving a sufficiently small product $\sigmatot r_p$. For example, Eq.~(\ref{eq:sail_sigma_req}) gives
\(\sigmatot\simeq2.3~\gmsq\) for \(\vinf\simeq155~\kms\),
\(\rp=0.05~\AU\), and \(\chis=1\). Relaxing the perihelion to
\(\rp=0.15~\AU\) would require \(\sigmatot\simeq0.75~\gmsq\) to reach the same \(\vinf\), whereas going to only a few solar radii relaxes the areal-density requirement but tightens thermal-survivability, attitude-control, and optical-degradation margins. These results push the feasibility question into coupled materials, thermal, and large-area deployment qualification at mission scale. This assessment is consistent with the conclusion here: the 25--40~yr-class sail regime is plausibly within reach if scale-up and thermal-property stability are matured, whereas the sub-20~yr sail-only regime requires pushing multiple coupled technology dimensions simultaneously (ultra-low $\sigmatot$, very deep $\rp$, and survivability at mission scale).

Importantly, this maturity gap is not a physics limit: it is a program-and-demonstration limit. A focused late-2020s/early-2030s development that couples (i) large-area deployment validation, (ii) deep-perihelion optical-property stability tests, and (iii) integrated areal-density demonstrations at the $10^4$--$10^5~\mathrm{m^2}$ scale could credibly raise the SGL-class sail system TRL into the mission-start window, particularly for the 25--40 yr-class access regime.

\section{Nuclear electric propulsion (NEP) to the SGL}
\label{sec:NEP}

\subsection{Why NEP is attractive for SGL even if it is not the fastest}

At $650$--$900~\AU$, solar flux is negligible; mission power must be radioisotope or fission.
Solar electric propulsion (SEP) is therefore not an enabling power source for SGL operations at these distances, even though it is attractive in the inner solar system. NEP is especially attractive for SGL because it couples a long-lived non-solar electrical power source to high-$I_{\rm sp}$ propulsion, enabling (i) large outbound $\Delta v$ when operated for years and (ii) sustained electrical power for communications, attitude control, and image-plane mobility during SGL operations. (SEP can provide high-$I_{\rm sp}$ propulsion in the inner solar system but is not an enabling power source at $z\gtrsim 650~\AU$; NTP can provide high thrust for injection but does not provide long-lived electric power for the focal-region phase.)  For the high-voltage, high-power EP cases considered here, electromagnetic interference (EMI) and   electromagnetic compatibility (EMC) become a
system-level verification constraint that couples PMAD, PPU, plume interactions, avionics compatibility, and autonomous fault management.

\subsection{Baseline NEP-only access: what is achievable with constant-power models?}

\begin{figure}[t]
\includegraphics[width=0.65\linewidth]{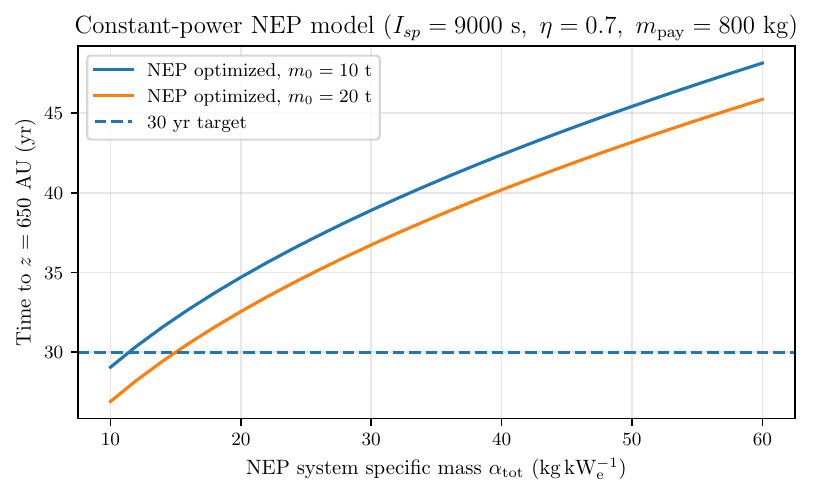}
\caption{Optimized reported outbound-leg time \(\trep\) to \(z=650~\AU\) versus integrated NEP stage specific mass \(\alphatot\) for \(\Isp=9000~\mathrm{s}\), \(\eta=0.7\), and \(\mpay=800~\mathrm{kg}\). Each point is optimized over propellant fraction under the constant-\(\Pe\) lower-bound model.}
\label{fig:nepalpha}
\end{figure}

\begin{figure}[t]
\includegraphics[width=0.65\linewidth]{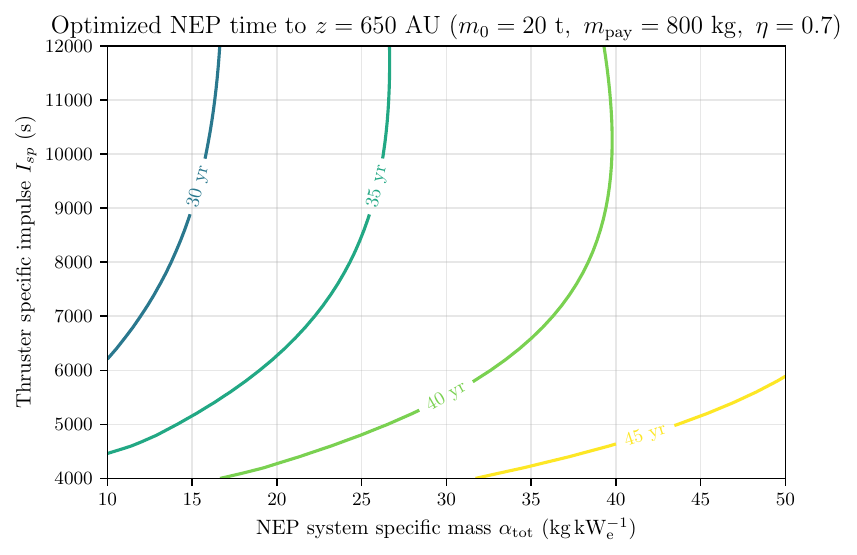}
\caption{Optimized reported outbound-leg time \(\trep\) to \(z=650~\AU\) for a \(m_0=20~\mathrm{t}\) wet-mass spacecraft with \(\mpay=800~\mathrm{kg}\) and \(\eta=0.7\). Contours show the coupled lower-bound requirements on \(\Isp\) and \(\alphatot\) under the constant-\(\Pe\) model.}
\label{fig:nepcontour}
\end{figure}

We use Eqs.~(\ref{eq:nep_power_balance})--(\ref{eq:wet_mass}) and Appendix~\ref{app:nep_time_model} to compute the reported outbound-leg time \(\trep\) to \(z=650~\AU\) as a function of \(\alphatot\), \(\Isp\), and propellant fraction, for representative wet masses \(m_0=10\) and \(20~\mathrm{t}\) with \(\mpay=800~\mathrm{kg}\) and \(\eta=0.7\). Each point is optimized over propellant fraction. Figure~\ref{fig:nepalpha}
shows \(\trep\) versus \(\alphatot\) for \(\Isp=9000~\mathrm{s}\); Fig.~\ref{fig:nepcontour} shows \(\trep\) contours over
\((\alphatot,\Isp)\) for \(m_0=20~\mathrm{t}\). Table~\ref{tab:nep_points} lists optimized design points for
\(m_0=20~\mathrm{t}\). Because \(\alphatot\) is a stage-level closure parameter, applying conventional mass growth or contingency to a current-best-estimate point design can be interpreted, to first order, as an upward shift in effective \(\alphatot\), moving the operating point along the same \(\trep\) contours in Fig.~\ref{fig:nepcontour} and
Table~\ref{tab:nep_points}.

These NEP-only values are therefore best interpreted as post-escape lower bounds. A physically closed trajectory must specify how the vehicle reaches the modeled initial state, because the initial thrust acceleration in the optimized cases is comparable to or smaller than solar gravity inside several AU [Eq.~(\ref{eq:nep_gravity_crossover})]. A full low-thrust solution would replace the rectilinear \(v_0\simeq 0\) initialization with an optimal spiral/escape and steering history; that elapsed time is not included in Table~\ref{tab:nep_points}.

\begin{table}[t]
\caption{Optimized NEP design points to reach \(z=650~\AU\) in the 1D constant-power model (\(\Isp=9000~\mathrm{s}\), \(\eta=0.7\), \(\mpay=800~\mathrm{kg}\), wet mass \(m_0=20~\mathrm{t}\)). \(\alphatot\) is the integrated stage-level specific mass (power + propulsion). The reported transfer time is the outbound-leg time \(\trep\) defined in Sec.~\ref{sec:tof_scope}; it excludes architecture-dependent launch, targeting,
and injection overhead.}
\label{tab:nep_points}
%\begin{ruledtabular}
\begin{tabular}{cccccccc}
\hline
\(\alphatot\) (\(\kgperkWe\)) &
\(\Pe\) (\(\kWe\)) &
\(m_0/\mf\) &
\(\mprop\) (t) &
\(v_f\) (\(\kms\)) &
\(v_f\) (\(\AUyr\)) &
\(t_b\) (\(\yr\)) &
\(\trep\) (\(\yr\))\\
\hline\hline
10 & 266 & 5.77 & 16.54 & 155 & 32.6 & 10.9 & 26.9\\
12 & 241 & 5.41 & 16.31 & 149 & 31.4 & 11.9 & 28.2\\
15 & 213 & 5.00 & 16.00 & 142 & 30.0 & 13.2 & 30.0\\
20 & 181 & 4.52 & 15.58 & 133 & 28.1 & 15.2 & 32.6\\
30 & 143 & 3.94 & 14.92 & 121 & 25.5 & 18.4 & 36.7\\
40 & 120 & 3.58 & 14.41 & 112 & 23.7 & 21.2 & 40.2\\
\hline
\end{tabular}
%\end{ruledtabular}
\end{table}

A key point for interpreting these NEP design points is that a large terminal speed does not by itself guarantee a short transfer. With constant electrical power the thrust is only a few newtons at hundreds of kWe, so the vehicle spends many years accelerating and covers only a modest fraction of the total distance during the burn. The remaining hundreds of AU are traversed during coast at the achieved terminal speed. Meeting a strict $\le 20\,\mathrm{yr}$ requirement without an injected $v_0$ therefore demands either (i) much higher thrust-to-mass (larger $\Pe$ at the same $m_0$, which in turn requires much lower $\alphatot$), or (ii) a hybrid architecture that supplies a large $v_0$ prior to initiating the NEP burn.

Several conclusions are robust:
\begin{itemize}
\item Even with aggressive \(\alphatot\sim10~\kgperkWe\), NEP-only \(\trep\) to $\SI{650}{\AU}$ is $\sim \SI{27}{yr}$ for a $20$~t class spacecraft, dominated by the cruise distance.
\item Achieving \(\lesssim 30~\yr\) requires
\(\alphatot\lesssim 15\)--\(20~\kgperkWe\) with
\(\Isp\gtrsim 8000~\mathrm{s}\).
\item High $I_{sp}$ is mandatory for large outbound $\Delta v$: without $I_{sp}\gtrsim \SI{5000}{s}$, mass ratios needed for $\Delta v\sim \SI{100}{\kms}$ become extreme (Fig.~\ref{fig:nepcontour}).
\end{itemize}

The technology implications of these parametric requirements (high-voltage EP operation, multi-year throughput and life qualification, and large-area heat rejection) are discussed explicitly in Secs.~\ref{sec:nep_closure} and \ref{sec:nep_readiness}.

\subsection{NEP system-closure checks: thrust-to-mass, total impulse, high-voltage operation, and heat rejection}
\label{sec:nep_closure}

The optimized times in Figs.~\ref{fig:nepalpha}--\ref{fig:nepcontour}
can be misread unless the implied thrust-to-mass and lifetime requirements
are made explicit. For constant electrical power,
\(\Thrust\propto\Pe/\Isp\), and the initial acceleration is
\begin{equation}
a_0 =
\frac{\Thrust}{m_0}
=
\frac{2\eta\Pe}{m_0g_0\Isp}.
\label{eq:nep_initial_accel}
\end{equation}
For the \(m_0=20~\mathrm{t}\) reference cases, typical optima require \(\Pe\simeq0.2\)--\(0.35~\MWe\) at
\(\Isp\simeq8000\)--\(9000~\mathrm{s}\), implying only few-newton thrust. For example, \(\Pe=300~\kWe\), \(\eta=0.7\), and
\(\Isp=9000~\mathrm{s}\) give \(\Thrust\simeq4.8~\mathrm{N}\) and \(a_0\simeq2.4\times10^{-4}~\mathrm{m\,s^{-2}}\), or
\(\sim7.6~\mathrm{km\,s^{-1}\,yr^{-1}}\) before mass depletion increases acceleration.

The second closure quantity is total impulse. For fixed \(\Isp\), the
integrated impulse delivered by the electric-propulsion system is
\begin{equation}
I_{\rm tot} =
\int \Thrust\,dt
\simeq \mprop g_0\Isp,
\label{eq:total_impulse}
\end{equation}
which depends primarily on propellant throughput rather than on thrust history. For the representative \(\mprop\simeq15\)--\(17~\mathrm{t}\) cases in Table~\ref{tab:nep_points} with \(\Isp=9000~\mathrm{s}\), this implies \(I_{\rm tot}\sim 1.3\)--\(1.5\times 10^{9}~\mathrm{N\,s}\). This requirement should be compared to demonstrated long-duration electric propulsion qualification at flight-relevant conditions: the SGL NEP problem is dominated by multi-year throughput and erosion management, not by achieving a particular \(\Isp\) at a single operating point.

The most relevant publicly documented long-duration benchmark for high-$\Isp$ electric propulsion is the NEXT ion-thruster long-duration test, which exceeded $4.2\times 10^4$~h of operation, processed $>736$~kg of xenon, and delivered $>2.8\times 10^{7}$~N\,s total impulse \cite{NEXTLDT2012}. Later status reporting documents $>5.0\times 10^4$~h operation, $\sim 902$~kg throughput, and $\sim 3.5\times 10^{7}$~N\,s total impulse \cite{NEXTLDT2015}. By comparison, the SGL NEP reference cases require $I_{\rm tot}\sim 10^{9}$~N\,s at the spacecraft level. This does not imply a single thruster must deliver $10^{9}$~N\,s: a clustered architecture trades total impulse across many strings. For example, a $\Pe\simeq 300~\mathrm{kW_e}$ stage implemented as $\sim 40$ strings at $\sim 7$--$8~\mathrm{kW}$ each would process $\sim 15$--$17~\mathrm{t}$ total propellant as $\sim 0.4$~t per thruster, comparable to demonstrated NEXT-class throughput on a per-thruster basis. The \emph{dominant} remaining gap for the SGL-fast cases is therefore not merely throughput, but simultaneously achieving (i) substantially higher exhaust voltage than NEXT-class operation and (ii) multi-string reliability, plume/EMI compatibility, and lifetime qualification at that voltage and duty cycle.

Third, high-\(\Isp\) operation implies high acceleration potentials for ion-class thrusters. A singly-charged ion accelerated through an electrostatic potential difference \(V\)
(i.e., the beam/acceleration voltage in an ion thruster) has exhaust speed \(v_e\simeq \sqrt{2qV/m_i}\), giving an order-of-magnitude relation
\begin{equation}
V \simeq \frac{m_i (g_0\Isp)^2}{2q}.
\end{equation}
For xenon, \(\Isp=9000~\mathrm{s}\) corresponds to \(V\sim 5~\mathrm{kV}\), while \(\Isp\simeq 12000~\mathrm{s}\) pushes \(V\) toward \(\sim 10~\mathrm{kV}\) (propellant- and charge-state-dependent). This places demanding requirements on power processing, insulation, contamination control, and grid/keeper erosion for decade-scale operation at hundreds of kilowatts.

Finally, fission-electric NEP at \(\Pe\sim 0.2\)--\(0.4~\mathrm{MW_e}\) is a thermal-management problem as much as a propulsion problem. Let \(\etapc\) be the reactor-to-electric conversion efficiency. The rejected heat is
\begin{equation}
Q_{\rm rej}=\Pe\left(\etapc^{-1}-1\right),
\label{eq:qrej}
\end{equation}
and the ideal radiator area is
\begin{equation}
A_{\rm rad}\simeq
230~\mathrm{m^2}
\left(\frac{\Pe}{300~\kWe}\right)
\left[\frac{(\etapc^{-1}-1)}{2.33}\right]
\left(\frac{0.85}{\epsilon}\right)
\left(\frac{500~\mathrm{K}}{T_{\rm rad}}\right)^4 .
\label{eq:arad_scaled}
\end{equation}
The corresponding contribution to stage specific mass is
\begin{equation}
\alpha_{\rm rad}\simeq
3.8~\kgperkWe
\left(\frac{\gammarad}{5~\mathrm{kg\,m^{-2}}}\right)
\left(\frac{A_{\rm rad}}{230~\mathrm{m^2}}\right)
\left(\frac{300~\kWe}{\Pe}\right).
\label{eq:alpha_rad}
\end{equation}
Therefore, a radiator subsystem alone can consume several \(\mathrm{kg\,kW_e^{-1}}\) of the \(\alphatot=10\)--\(20~\kgperkWe\) budget, before reactor, shield, conversion, PMAD, PPU, thrusters, harnessing, and redundancy are included.

\begin{table}[t]
\caption{Representative fast-NEP burden translated to a 40-string cluster
for an SGL transfer case with \(\Pe=300~\mathrm{kW_e}\), \(\eta=0.7\),
\(I_{sp}=9000~\mathrm{s}\), total propellant \(\mprop=15\)--\(17~\mathrm{t}\),
and full-power operation of \(\sim 9\)--\(10~\mathrm{yr}\).}
\label{tab:nep_cluster}
\centering
\small
\begin{tabular}{lcc}
\hline
Quantity & Spacecraft level & Per string (40 strings) \\
\hline\hline
Electrical power & \(300~\mathrm{kW_e}\) & \(7.5~\mathrm{kW_e}\) \\
Thrust & \(4.8~\mathrm{N}\) & \(0.12~\mathrm{N}\) \\
Propellant throughput & \(15\)--\(17~\mathrm{t}\) & \(0.38\)--\(0.43~\mathrm{t}\) \\
Total impulse & \((1.3\)--\(1.5)\times 10^9~\mathrm{N\,s}\) &
\((3.3\)--\(3.8)\times 10^7~\mathrm{N\,s}\) \\
Full-power duration & \(9\)--\(10~\mathrm{yr}\) & \(9\)--\(10~\mathrm{yr}\) \\
\hline
\end{tabular}
\end{table}

Table~\ref{tab:nep_cluster} helps separate two issues that are often conflated. At the spacecraft level, the SGL propulsion is an \(\mathcal{O}(10^9~\mathrm{N\,s})\) total-impulse problem. At the string level, however, a clustered architecture can place the throughput and total impulse per thruster in the vicinity of demonstrated NEXT-class values. The remaining gap is therefore not simply per-string throughput, but the simultaneous closure of high-voltage lifetime, multi-string reliability, EMI/EMC, plume compatibility, and large-scale thermal management.

\subsection{Can NEP reach the SGL in under 20 years? Requirements and the role of hybrid injection}

The baseline NEP-only results above assume that NEP begins from modest heliocentric excess speed, effectively $v_0\simeq 0$ in the long outbound leg. Under this assumption, a $\SI{20}{yr}$ transfer to $\SI{650}{\AU}$ is only possible if $\alphatot$ is far below values commonly assumed for near-term space fission power. For \(m_0=20~\mathrm{t}\), \(\mpay=800~\mathrm{kg}\),
\(\eta=0.7\), and \(\Isp=9000~\mathrm{s}\), the required
stage specific mass for \(\trep\le 20~\yr\) is approximately
\begin{equation}
\alphatot \lesssim 2.8~\kgperkWe
\qquad
(v_0\simeq 0,\ \Isp=9000~\mathrm{s}).
\label{eq:nep_alpha_20yr}
\end{equation}
If \(\Isp\) is increased to \(\sim 1.2\times10^4~\mathrm{s}\),
\(\trep\le20~\yr\) can be achieved at
\(\alphatot\sim3\)--\(4~\kgperkWe\) in this simplified model,
but this demands both extremely lightweight fission-electric power
and high-voltage, long-life electric thrusters.\footnote{The
\(\alphatot\lesssim3~\kgperkWe\) regime is used here as an
aspirational lower bound to quantify the thrust-to-mass requirement; it should not be interpreted as a claim of near-term achievability for a flight-ready fission-electric stage.}

\begin{figure}[t]
\includegraphics[width=0.65 \linewidth]{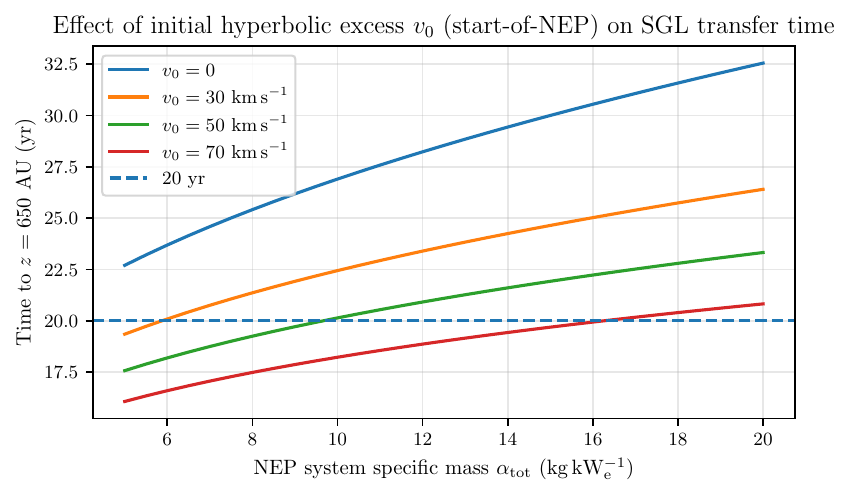}
\caption{Effect of injected hyperbolic excess speed \(v_0\), specified at the start of the modeled NEP outbound leg, on optimized reported time \(\trep\) to \(z=650~\AU\). Assumptions: \(m_0=20~\mathrm{t}\), \(\mpay=800~\mathrm{kg}\), \(\eta=0.7\), and \(\Isp=9000~\mathrm{s}\). A \(20~\yr\) lower-bound transfer is inaccessible for \(v_0\simeq0\) unless \(\alphatot\lesssim3~\kgperkWe\), but becomes possible for
\(\alphatot\sim10\)--\(15~\kgperkWe\) if an upstream injection architecture supplies \(v_0\simeq50\)--\(70~\kms\).}
\label{fig:nepv0}
\end{figure}

A more realistic pathway to \(\trep\le20~\yr\) with
\(\alphatot\sim10\)--\(15~\kgperkWe\) is a hybrid architecture in
which a high-thrust injection stage supplies a substantial initial
hyperbolic excess \(v_0\) before NEP cruise. Figure~\ref{fig:nepv0}
shows the sensitivity of optimized NEP \(\trep\) to \(v_0\). For
\(\Isp=9000~\mathrm{s}\), \(\trep\le20~\yr\) becomes possible at
\begin{equation}
\alphatot\lesssim 9.7~\kgperkWe
\quad \mathrm{for}\quad
v_0\gtrsim50~\kms,
\label{eq:hybrid_alpha_50kms}
\end{equation}
or at \(\alphatot\lesssim12.5~\kgperkWe\) for
\(v_0\gtrsim60~\kms\). A practical advantage of hybrid injection is that it reduces the required \emph{full-power} NEP operating duration to $t_b\approx 9$--10~yr (Table~\ref{tab:nep_hybrid}), relaxing lifetime and qualification demands relative to NEP-only cases where $t_b$ can exceed a decade and approach two decades (Table~\ref{tab:nep_points}).

\begin{table}[t]
\caption{Representative hybrid injection + NEP design points that approach or beat \(\trep=20~\yr\) in the 1D constant-power model. Assumptions: \(m_0=20~\mathrm{t}\) at the start of the NEP phase, payload \(\mpay=800~\mathrm{kg}\), \(\Isp=9000~\mathrm{s}\), and electrical efficiency \(\eta=0.7\). For each \((v_0,\alphatot)\) pair we sweep \(\Pe\) and report the \(\Pe\) that minimizes \(\trep\). The reported transfer time is the
outbound-leg time \(\trep\) defined in Sec.~\ref{sec:tof_scope}; it excludes the mass and elapsed-time cost of the upstream injection architecture that provides \(v_0\).}
\label{tab:nep_hybrid}
\begin{tabular}{ccccccccc}
\hline
\(v_0\) (\(\kms\)) &
\(\alphatot\) (\(\kgperkWe\)) &
\(\Pe\) (\(\kWe\)) &
\(\mf\) (t) &
\(\mprop\) (t) &
\(m_0/\mf\) &
\(t_b\) (\(\yr\)) &
\(v_f\) (\(\kms\)) &
\(\trep\) (\(\yr\))\\
\hline\hline
50 & 9.7  & 323 & 3.93 & 16.07 & 5.10 &  8.77 & 194 & 20.02 \\
60 & 12.5 & 304 & 4.60 & 15.40 & 4.35 &  8.93 & 190 & 20.01 \\
70 & 15.0 & 274 & 4.91 & 15.09 & 4.07 &  9.71 & 194 & 19.69 \\
\hline
\end{tabular}
\end{table}

%\paragraph{Hybrid cases as requirement bounds.}
In the hybrid trades, the injected speed \(v_0\) is treated as a boundary condition at the start of the NEP phase rather than as a closed design variable. Accordingly, Table~\ref{tab:nep_hybrid} states the required performance of an upstream injection architecture, not the existence of a qualified one. In first-order form, any such injection stage must satisfy
\begin{equation}
\Big(\frac{m_{\rm inj,0}}{m_0}\Big)_{\rm ideal}
=
\exp \Big(\frac{\Delta v_{\rm inj}}{g_0 I_{\rm sp,inj}}\Big),
\label{eq:inj_ratio}
\end{equation}
with additional penalties from inert mass, thermal protection, cryogenic or propellant
management, and perihelion operations. Hybrid cases that approach \(\sim 20~\yr\)
should therefore be read as requirement boundaries on the injection stage, not as
end-to-end mass-closed mission solutions.

Figures~\ref{fig:nepalpha}--\ref{fig:nepv0} show an important but easily-misinterpreted point: even when the optimized final hyperbolic speed approaches the $\SI{154}{\kms}$ mean-speed requirement, the transfer time remains longer than the naive ``$z/v$'' estimate because a constant-thrust vehicle spends a multi-year burn at substantially lower speed. Therefore, a sub-$\SI{20}{yr}$ SGL transfer cannot be judged from $v_f$ alone; it requires both (i) sufficiently high final speed \emph{and} (ii) sufficiently high early acceleration.

A compact way to expose the requirement is to combine
Eq.~(\ref{eq:nep_thrust_mdot}) with \(a=\Thrust/m\). At the start of the burn,
\begin{equation}
a_0 =
\frac{\Thrust}{m_0}
=
\frac{2\eta\Pe}{m_0g_0\Isp},
\label{eq:hybrid_initial_accel}
\end{equation}
and the propellant consumption rate is
\begin{equation}
\dot{m} =
\frac{2\eta\Pe}{g_0^2\Isp^2}.
\label{eq:hybrid_mdot}
\end{equation}
For the $\SI{20}{t}$ reference spacecraft, achieving $\mathcal{O}(10^5~\mathrm{m\,s^{-1}})$ class total $\Delta v$ within a decade-scale burn generically pushes $\Pe$ to the few$\times 10^2~\kWe$ regime and demands a very large propellant throughput. This drives (a) NEP specific mass $\alphatot$, (b) thruster life and erosion limits, and (c) heat-rejection deployability as first-order feasibility constraints rather than second-order implementation details.

A practical way to interpret Table~\ref{tab:nep_hybrid} is to translate the ``few\(\times10^2~\kWe\)'' requirement into thrust level, acceleration, and propellant logistics. For \(\Pe\simeq300~\kWe\), \(\eta=0.7\), and
\(\Isp=9000~\mathrm{s}\), the ideal thrust is
\(\Thrust=2\eta\Pe/(g_0\Isp)\simeq4.8~\mathrm{N}\), implying an initial acceleration of only \(a_0\simeq\Thrust/m_0\simeq
2.4\times10^{-4}~\mathrm{m\,s^{-2}}\) for \(m_0=20~\mathrm{t}\). The corresponding propellant mass flow from Eq.~(\ref{eq:hybrid_mdot}) is
\(\dot{m}\simeq5\times10^{-5}~\mathrm{kg\,s^{-1}}\), i.e.,
\(\sim4\)--\(5~\mathrm{kg\,day^{-1}}\) or
\(\sim1.5\)--\(2~\mathrm{t\,yr^{-1}}\). Therefore, the sub-20~yr hybrid cases in Table~\ref{tab:nep_hybrid} implicitly require multi-year continuous operation with total propellant throughput at the $\sim 15$--$16$~t level, while maintaining high-voltage performance and life against erosion. This is the dominant NEP realism discriminator: the mission is not merely ``high-$\Isp$,'' but ``high-$\Isp$ at high power for a decade-scale integrated burn with very large throughput,'' which couples directly to thruster lifetime qualification, cathode/keeper wear-out, plume interaction with deployable radiators, and tankage/propellant feed system design.

\begin{table}[t]
\caption{Sensitivity of optimized NEP reported outbound-leg time to achievable \(\Isp\) and injection speed \(v_0\) for a representative system-specific-mass case
\((\alphatot=15~\kgperkWe,\; m_0=20~\mathrm{t},\;
\mpay=800~\mathrm{kg},\; \eta=0.7,\; z=650~\AU)\).
\(P_e^\star\) is the electrical power that minimizes \(\trep\) under the mass-closure model \(m_{\rm dry}=\mpay+\alphatot\Pe\).}
\label{tab:nep_sens}
%\begin{ruledtabular}
\begin{tabular}{c c c c c}
\hline
\(\Isp\) (s) &
\(v_0\) (\(\kms\)) &
\(\Pe^\star\) (\(\kWe\)) &
\(t_b^\star\) (\(\yr\)) &
\(\trep\) (\(\yr\))\\
\hline\hline
4000 & 0  &  65  &  9.7 & 39.1 \\
4000 & 70 &  93  &  6.7 & 22.1 \\
6000 & 0  & 121  & 11.3 & 33.2 \\
6000 & 70 & 163  &  8.0 & 20.5 \\
8000 & 0  & 182  & 12.6 & 30.7 \\
8000 & 70 & 237  &  9.2 & 19.9 \\
9000 & 0  & 213  & 13.2 & 30.0 \\
9000 & 70 & 274  &  9.7 & 19.7 \\
\hline
\end{tabular}
%\end{ruledtabular}
\end{table}

Although the reported outbound-leg time to \(\sim650~\AU\) is
multi-decade in many cases, the fission-electric system
is required to operate at full power primarily during the NEP burn ($t_b$; Tables~\ref{tab:nep_points}--\ref{tab:nep_hybrid}). After burnout the vehicle coasts for hundreds of AU, during which the reactor/EP system could in principle be throttled or placed in a low-power state (subject to restart and cruise-power requirements). For technology maturation and mission assurance it is therefore useful to specify both (i) full-power operating years (set by $t_b$ and the science phase) and (ii) total calendar years to end-of-mission.

Table~\ref{tab:nep_sens} makes the sub-$\SI{20}{yr}$ conclusion precise: for \(\alphatot\sim10\)--\(15~\kgperkWe\), a $\le\SI{20}{yr}$ transfer requires both a large injection speed ($v_0\sim$ few$\times 10~\kms$) and an electric propulsion system capable of sustained multi-year operation at $I_{sp}\gtrsim \SI{6000}{-}\SI{8000}{s}$ with very high propellant throughput. In the absence of the high-thrust injection ($v_0\simeq 0$), NEP-only transfers remain multi-decade even for optimistic $I_{sp}$.

\subsection{NEP technology readiness and programmatic gates for a 2035--2040 SGL start}
\label{sec:nep_readiness}

Technology readiness for NEP must be assessed at the integrated stage level, not at the component level.
For SGL-relevant missions, the required capability is not ``an electric thruster'' or ``a space reactor'' in isolation, but a coupled flight system: reactor+shield, power conversion, PMAD, primary heat rejection, and a redundant EP string that can operate for multi-year durations with high propellant throughput. In NASA TRL terms, this corresponds to demonstrating an integrated subsystem in a relevant environment (TRL~6) before committing a flagship-class science mission to it.

NASA has published dedicated maturation plans for MW-class NEP that explicitly treat the reactor/conversion/ thermal/PMAD/EP stack as an integrated system, and identify radiator packaging and long-duration EP operation as first-order constraints for transport-class missions \cite{MartinTMP2022,MasonTM2022}. Consistent with that framing, NASA TechPort reports the (now completed) NEP Technology Maturation project with current TRL~3 and an end-target of TRL~5, underscoring that publicly documented NEP efforts have not yet closed to an integrated flight stage at $\Pe\gtrsim 100~\mathrm{kW_e}$. In parallel, NASA has investigated modular assembled radiators (MARVL) as a mitigation for the deployable area and packaging constraints that dominate megawatt-class concepts \cite{MARVL}. Accordingly, for a 2035--2040 SGL start, the central readiness question is not whether individual components exist, but whether an end-to-end stage (power source, conversion, heat rejection, PMAD, and multi-string EP) can be demonstrated at relevant scale and duration early enough to support mission Phase~B.

NASA's public NEP technology maturation activities have historically targeted raising key elements toward mid-TRL, but that does not equate to an SGL-ready stage.
A realistic 2035--2040 SGL program that depends on \(\Pe\sim 0.2\)--\(0.4~\mathrm{MW_e}\) NEP must therefore either (i) inherit a prior flight demonstration of an integrated fission-electric bus at \(\gtrsim 100~\mathrm{kW_e}\), or (ii) explicitly include such a demonstration as a schedule-critical precursor within the overall program. In the absence of an integrated demonstration, the schedule risk is dominated by nuclear safety approval, end-to-end thermal testing, EMI/EMC interactions, plume/radiator compatibility, and long-duration operations qualification rather than by analytic performance.

A practical set of programmatic gates consistent with a 2035--2040 launch is:
\begin{enumerate}
\item By \(\sim 2028\): downselect to a specific EP class (ion vs Hall vs alternative) and propellant family consistent with the required \(\Isp\) and throughput; complete life test planning and begin qualification tests at flight-like voltages and power densities.
\item By \(\sim 2030\): demonstrate reactor-representative power conversion and PMAD at relevant power with end-to-end load following, fault management, and radiation-hard control electronics; demonstrate radiator deployability at relevant area and packaging density, including damage tolerance and fluid-loop/heatpipe robustness.
\item By \(\sim 2031\)--2032: complete an integrated ``nuclear-simulator'' NEP stage test (reactor simulator + conversion + radiators + multi-string EP) to retire system-level coupling risks (thermal dynamics, plume impingement, EMI/EMC, control authority, and safe-mode behaviors).
\item By \(\sim 2033\): if NEP is a mission-critical transportation-time driver, fly an in-space integrated demonstration (cis-lunar or heliocentric) long enough to anchor life predictions and close the verification argument at the system level.
\end{enumerate}
Without these gates, a 2035 launch is still possible, but NEP should be treated as an operations enabler (power and fine-control at the focal region) rather than as the primary transportation system expected to deliver \(\trep \lesssim 20~\yr\).

\subsection{Propellant logistics and cost realism}

Fast NEP transfers require \(\mathcal{O}(10~\mathrm{t})\) of propellant for an \(\mathcal{O}(10\)--\(20~\mathrm{t})\) spacecraft when \(\Isp\sim 8000\)--\(9000~\mathrm{s}\) (Table~\ref{tab:nep_points}). The storage-volume scale is
\begin{equation}
V_{\rm prop}\simeq
10~\mathrm{m^3}
\left(\frac{\mprop}{16~\mathrm{t}}\right)
\left(\frac{1.6~\mathrm{t\,m^{-3}}}{\rho_{\rm store}}\right),
\label{eq:prop_volume}
\end{equation}
where \(\rho_{\rm store}\) is the effective flight storage density including usable-fluid fraction and tankage. Xenon minimizes some thruster-performance risk but can be supply- and cost-constrained at the multi-ton scale; krypton or argon may reduce propellant procurement burden at the cost of thruster redesign and possibly lower efficiency, while iodine offers high storage density but requires separate high-power lifetime closure. These constraints do not invalidate NEP, but they make the SGL-fast NEP option a large-spacecraft logistics problem rather than a smallsat-class extension.

\section{Nuclear thermal propulsion (NTP) and solar Oberth injection}
\label{sec:NTP}

\subsection{Energetic limitations of NTP as a standalone SGL solution}

With $I_{sp}\sim \SI{900}{s}$, NTP can provide high thrust but is constrained by the rocket equation. Achieving $\Delta v\sim \SI{100}{\kms}$ directly is infeasible at reasonable mass ratios; therefore, NTP should not be viewed as a standalone means to reach the SGL rapidly. Its most compelling role is as a high-thrust injection stage executed deep in the solar gravitational well (Oberth maneuver), where Eq.~(\ref{eq:oberth}) yields large $v_\infty$ for modest $\Delta v$.

\subsection{Oberth requirements and thermal/cryogenic realism}

The Oberth relations used here describe an ideal impulsive burn in heliocentric vacuum at perihelion; they are not an atmospheric launch model. The purpose is to translate the injected-speed boundary condition \(v_0\) required by the hybrid NEP trades into an ideal high-thrust stage burden.

For a parabolic perihelion pass at radius \(\rp\), \(v_p=\sqrt{2\mu_\odot/\rp}\). An impulsive prograde burn that produces post-burn hyperbolic excess \(\vinf\) requires
\begin{equation}
\Delta v_{\rm O} =
\sqrt{v_p^2+\vinf^2}-v_p .
\label{eq:oberth_dv_required}
\end{equation}
Table~\ref{tab:oberth_injection} translates the hybrid-relevant \(v_0=50\)--\(70~\kms\) range into ideal burn requirements for \({\Isp}_{\rm inj}=900~\mathrm{s}\). The values are ideal: they exclude inert mass, thermal protection, boiloff or cryogenic conditioning, finite-burn losses, guidance margin, and the elapsed time to target perihelion.

\begin{table}[t]
\caption{Ideal solar-Oberth injection burden for an \({\Isp}_{\rm inj}=900~\mathrm{s}\) high-thrust stage. The mass ratio is \(\exp[\Delta v_{\rm O}/(g_0{\Isp}_{\rm inj})]\).}
\label{tab:oberth_injection}
\centering
\small
\begin{tabular}{ccccc}
\hline\hline
Target \(\vinf\) &
\multicolumn{2}{c}{\(\rp=0.02~\AU\)} &
\multicolumn{2}{c}{\(\rp=0.05~\AU\)} \\
\cline{2-5}
(\(\kms\)) &
\(\Delta v_{\rm O}\) (\(\kms\)) & \(m_0/m_f\) &
\(\Delta v_{\rm O}\) (\(\kms\)) & \(m_0/m_f\) \\
\hline
50  & 4.2  & 1.60 & 6.5  & 2.09 \\
60  & 6.0  & 1.97 & 9.3  & 2.88 \\
70  & 8.1  & 2.51 & 12.6 & 4.16 \\
95  & 14.8 & 5.34 & 22.6 & 12.9 \\
155 & 37.9 & 73.4 & 55.6 & \(5.4\times 10^2\) \\
\hline\hline
\end{tabular}
\end{table}

The table reinforces the architectural conclusion: NTP or another high-thrust stage is most credible as a \(50\)--\(70~\kms\) injection enabler for hybrid NEP trajectories, not as a standalone route to \(v_\infty\sim 155~\kms\) at useful delivered mass.

Recent space nuclear propulsion activity includes NASA investments in nuclear propulsion concepts and the DARPA/NASA DRACO nuclear thermal propulsion effort, whose public DARPA program page now lists the effort as complete and no longer maintained \cite{NASASNP,DARPA_DRACO}. The key engineering point for SGL is that NTP maturity should be evaluated at the integrated stage level (reactor + turbomachinery + cryogenic storage + thermal protection + guidance at perihelion), not solely at the fuel-element level.

\section{Technology readiness and roadmap for a 2035--2040 mission start}
\label{sec:TRL}

Technology readiness levels (TRLs) are used here in the NASA sense: TRL~6 corresponds to a system or subsystem demonstration in a relevant environment, while TRL~7 and above imply space demonstration and flight qualification. For integrated nuclear propulsion, the distinction between component TRL and end-to-end mission capability is crucial: high-power thrusters and reactor components can be advanced in isolation, yet the coupled system (reactor + conversion + PMAD + thermal management + EP string) remains low-TRL until demonstrated as an integrated stage. Programmatically, this integration step also triggers long-lead activities (safety analysis, test infrastructure, nuclear launch approval), which must be treated as schedule-critical path items rather than ``paperwork.''

A recurring failure mode in nuclear propulsion advocacy is to quote high TRLs for individual components (e.g., fuel forms, turbines, thrusters, heatpipes) and then implicitly treat the integrated stage as equally mature. For the SGL, that inference is invalid: the coupled verification problem (thermal control with large deployables, multi-string EP plume interactions, EMI/EMC across high-voltage PMAD, autonomous fault management, and nuclear launch approval) is what drives schedule and cost. Accordingly, the TRL values quoted in this section distinguish component TRL from stage TRL and treat integrated demonstration as the pacing requirement for a 2035--2040 start.

Cost realism: any architecture requiring a new space reactor plus high-power EP strings should be treated as flagship-class in both development scope and risk posture, unless a separate, sustained technology program retires reactor+PMAD+radiator+EP integration before Phase~B.

\subsection{Technology readiness levels (TRL) as a discipline}

NASA's TRL definitions provide a standardized measure of maturity from TRL~1 (basic principles observed) to TRL~9 (flight-proven through successful mission operations) \cite{NASA_TRL}. For propulsion selection, it is essential to distinguish (i) component TRL (e.g., a thruster demonstrated in a vacuum chamber) from (ii) system TRL (integrated propulsion stage demonstrated in a relevant environment), because SGL access depends on the latter.

\subsection{TRL snapshot for SGL-relevant propulsion elements}

Table~\ref{tab:trl} summarizes an indicative TRL snapshot.
Solar-sail subsystem TRLs are reported directly in the SGL mission-architecture study; other values are engineering estimates based on publicly reported maturity and should be treated as approximate.

Public programmatics introduce material schedule risk for NEP. As of early 2026, NASA TechPort lists the NEP Technology Maturation project (Project~158369) \cite{techport_nep_158369} as completed with $\mathrm{TRL}_{\rm current}=3$ (target $\mathrm{TRL}_{\rm end}=5$) and indicates close-out actions tied to the FY2026 budget profile. Therefore, a 2035--2040 SGL launch that relies on an integrated $0.1$--$0.3~{\rm MW_e}$ fission-electric stage would require either (i) reconstitution of a dedicated NEP maturation program with stable funding through qualification, or (ii) leveraging non-SNP nuclear power developments (e.g., surface power, DoD/DOE microreactors) to retire key risks in conversion, thermal rejection, and nuclear flight safety.

\begin{table}[t]
\caption{Indicative TRL snapshot for propulsion elements, emphasizing scale-to-mission effects. TRL definitions follow NASA guidance \cite{NASA_TRL}. ``SGL-class'' TRL is assessed at the scale and environment implied by \(z\simeq 650\)--\(900~\AU\) and (where relevant) deep-perihelion operation.}
\label{tab:trl}
\begin{tabular}{llcc}
\hline
Element (SGL-class) & Evidence/anchor & TRL (now) & TRL needed \\
 & & & for 2035--\\
  & & & 2040 start\\
\hline\hline
Solar sail deployment (subscale) & Flight demos at \(\sim\)10~m scale; structural deployment  & 6--8 & 8--9 \\
& heritage; limited area \cite{IKAROS,ACS3} & &  \\

Solar sail scale-up (\(\gtrsim 10^4\)--\(10^5~\mathrm{m^2}\)) & No flight demonstration at required area; packaging & 3--4 & 6--7 \\
& and dynamics scale nonlinearly &  &  \\

Solar sail deep-perihelion material system  & Survivability and optical-property stability at \(\gtrsim 700\)--& 2--3 & 6--7 \\
(\(\rp\lesssim 0.05~\AU\)) &\(900~\mathrm{K}\) not yet proven at scale &   &   \\

High-power EP (\(\sim 10\)--\(20~\mathrm{kW}\)) & Flight-proven Hall systems; gateway-class development  & 7--9 & 8--9 \\
 & heritage &   &   \\

High-power EP (\(\sim 100~\mathrm{kW}\) class) & Ground demonstrations; qualification/life remains  & 4--5 & 6--7 \\
 & pacing; cluster-level verification needed &   &  \\

High-\(\Isp\) EP at \(\gtrsim 100~\mathrm{kW}\) and \(\Isp\gtrsim 6000~\mathrm{s}\) & High-voltage lifetime/erosion at scale remains a key gap & 3--4 & 6--7 \\

Space fission power (10--40~kWe class) & Active development; relevant to near-term space fission  & 3--5 & 6--7 \\
& but not yet an NEP stage &   &  \\

Integrated NEP stage (100--300~kWe class) & Conceptual designs and element maturation; integrated & 2--3 & 6 \\
& stage demonstration not yet achieved &   &  \\

NTP injection stage (Oberth-capable) & Historical ground-test heritage; modern integrated stage  & 3--4 & 6 \\
& maturity dominated by fuel+CFM+TPS+operations &  &  \\
\hline
\end{tabular}
\end{table}

\subsection{Cost and programmatic considerations beyond TRL}
\label{sec:cost}

TRL captures technical maturity but does not by itself capture the dominant cost and schedule drivers for SGL
transportation (see Table~\ref{tab:costdrivers}). For this mission class, the primary discriminator among propulsion families is the amount of \emph{mission-unique non-recurring engineering} (NRE) required to reach an integrated, flight-qualifiable system,
together with the associated ground infrastructure and approval processes (especially for nuclear systems).

\paragraph{Deep-perihelion solar sailing.}
A sail-first transportation approach avoids nuclear flight approval and reactor ground-test infrastructure,
which can reduce programmatic cost and regulatory schedule risk. Its dominant mission-unique NRE is instead driven by (i) qualification of sail materials/coatings and optical-property stability in the $\sim 700$--$1000~\mathrm{K}$ regime, (ii) large-area deployment dynamics and metrology at the $10^{4}$--$10^{5}~\mathrm{m^2}$ scale, and (iii) validation of survivability and controllability at deep perihelion. Accordingly, sail-first architectures tend to be ``materials-and-deployment program'' limited rather than ``nuclear program'' limited.

\paragraph{Fission NEP.}
For NEP, the cost drivers are dominated by integrated-stage development and qualification: reactor+shielding,
power conversion, PMAD, large-area heat rejection, and multi-string EP life testing. Even when component demonstrations exist, end-to-end stage verification is typically the pacing item for cost and schedule. Public programmatics for NEP maturation can be tracked via NASA TechPort project reporting (e.g., Project~158369) \cite{techport_nep_158369}. Importantly, NEP is a crosscutting capability with potential users beyond SGL (cargo transport, high-power deep-space platforms), so the \emph{incremental} cost to an SGL mission depends strongly on whether NEP stage demonstrations are funded and flown for other purposes prior to Phase~B.

\paragraph{NTP/deep-solar Oberth injection.}
NTP-driven injection introduces additional programmatic complexity: reactor propulsion hardware, cryogenic
hydrogen storage and boil-off control, and (if used for deep-solar Oberth) perihelion thermal protection and operations. Recent program status should be treated as a schedule risk factor; for example, NASA TechPort reporting for DRACO indicates a stop-work memo dated 2~April~2025 \cite{techport_draco_105665}, implying that a 2035--2040 SGL architecture cannot assume availability of a flight-qualified NTP stage without a separately funded restart path.

\paragraph{Hybrid injection + NEP.}

Hybrid architectures can reduce \(\trep\) at realistic NEP specific mass, but they accumulate cost drivers from
both the high-thrust injection system and the NEP cruise stage, and they add integration/verification complexity
(e.g., deep-perihelion operations plus multi-year high-power EP plus large deployable thermal hardware).

\begin{table}[t]
\caption{Qualitative comparison of major mission-unique development and programmatic cost drivers for candidate SGL transportation families. ``Synergy'' indicates the degree to which costs can plausibly be amortized by non-SGL users (cargo transport, surface power, national programs).}
\label{tab:costdrivers}
\begin{tabular}{llll}
\hline\hline
Architecture & Dominant mission-unique NRE cost drivers & Major pacing items & Synergy \\
\hline
Deep-perihelion sail & High-$T$ sail materials/coatings; $10^{4}$--$10^{5}$ m$^2$  & Thermal qualification; large-scale  & Medium \\
& deployment, dynamics, metrology &  deployment tests &  \\[2pt]

Fission NEP & Integrated stage: reactor+conversion+ & Nuclear approval; long-duration EP  & High \\
& PMAD+radiators+EP string lifetime &  + thermal system verification &  \\[2pt]

NTP Oberth injection & Reactor propulsion stage; cryogenic H$_2$  & Nuclear qualification + genic/ & Medium \\
 &  storage; perihelion TPS/operations & perihelion ops validation &  \\[2pt]

Hybrid & Adds both injection and NEP stage costs; & Combined verification campaign; & Medium--High \\
(injection+NEP) & system integration complexity & long-lead nuclear + operations &  \\
\hline\hline
\end{tabular}
\end{table}

\subsection{Roadmap options consistent with a 2035--2040 start}

Two propulsion roadmaps appear most realistic for a 2035--2040 launch, depending on whether the mission driver is minimum \(\trep\) or maximum observatory capability.

\paragraph{Option A: Sail-first transportation, nuclear power for operations.}
Prioritize a close-perihelion solar sail to achieve $v_\infty\gtrsim 20~\AUyr$ with small-to-moderate payloads, and carry a radioisotope or small fission system to support operations at the SGL. This option is paced primarily by deep-perihelion sail material qualification (TRL~2$\rightarrow$6/7) and large-scale deployment.

\paragraph{Option B: Hybrid injection + NEP for cruise and operations.}
Use a high-thrust injection (deep solar Oberth) to achieve $v_0\gtrsim \SI{50}{\kms}$, then use a $\sim\SI{0.2}{\MWe}$--$\SI{0.5}{\MWe}$ NEP stage to add tens to hundreds of \(\kms\) over multi-year burns and to power a capable observatory at $650$--$900~\AU$. This option is paced by integrated NEP stage demonstration (system TRL~2/3$\rightarrow$6) and by the feasibility of a deep-perihelion injection stage that can survive and operate at $r_p\sim 0.02~\AU$.

\section{Conclusions}
\label{sec:concl}

For a 2035--2040 SGL mission start, propulsion selection is governed by both time-to-first-science and the delivered mass and electrical power available at \(z\simeq650\)--\(900~\AU\). A \(20~\yr\) arrival at \(650~\AU\) requires a mean radial speed of \(32.5~\AUyr\), or about \(154~\kms\), even before architecture-dependent injection and steering penalties are included. All transfer times quoted in this paper are the reported lower-bound outbound-leg times \(\trep\) defined in Sec.~\ref{sec:tof_scope}; they are not end-to-end mission-elapsed times.

The propulsion trade leads to three robust results. First, close-perihelion solar sailing remains the most schedule-credible non-nuclear route to very high \(\vinf\), but sub-\(20~\yr\) sail-only access lies in an extreme corner of \((\sigmatot,\rp)\) space that requires both deep-perihelion survivability and mission-scale ultra-low areal density. Second, NEP is attractive because it couples transportation to long-lived focal-region power and observatory control, but NEP-only transportation of the \(m_0=20~\mathrm{t}\), \(\mpay=800~\mathrm{kg}\) reference spacecraft remains \(\sim 27\)--\(33~\yr\) for \(\alphatot\simeq 10\)--\(20~\kgperkWe\). Third, \(\trep\lesssim 20~\yr\) becomes plausible only in hybrid architectures that combine a large injected \(v_0\sim 50\)--\(70~\kms\) with high-\(\Isp\), long-life EP, and large-scale thermal closure.

The programmatic implication is conditional. If the dominant objective is earliest credible access with a lightweight observatory, sail-first remains the most schedule-aligned architecture, paced primarily by deep-perihelion materials and large-area deployment qualification. If the dominant objective is a more capable observatory with larger delivered power and greater on-station maneuvering authority, hybrid injection+\;NEP is the preferred end-state, but only if an integrated \(\Pe\sim 0.2\)--\(0.4~\mathrm{MW_e}\) NEP stage and a credible
high-thrust injection precursor are demonstrated early enough to support mission commitment. In either case, the critical path is integrated system demonstration rather than any single component technology.

Finally, regardless of the transportation method, SGL operations at \(z\gtrsim 650~\AU\) require radioisotope or fission power; for sail-first architectures, that power-system mass is a first-order contributor to \(\sigmatot\) and therefore to achievable \(\vinf\). Future work must close the gap between the lower-bound envelopes reported here and mission feasibility by adding optimal-control trajectory design, realistic injection staging, steering losses, tankage and radiator packaging, end-to-end mass/power closure, and explicit reliability modeling for decade-scale power and propulsion operation.

%======= ACKNOWLEDGMENTS =============

\section*{Acknowledgments}
We are grateful to Louis D. Friedman and John O. Elliott for insightful comments and suggestions during the preparation of this manuscript. The work described here was carried out at the Jet Propulsion Laboratory, California Institute of Technology,
Pasadena, California, under a contract with the National Aeronautics and Space Administration. 
%\textcopyright\ 2026 California Institute of Technology. Government sponsorship acknowledged.

\appendix

\section{Analytic time-to-distance model for constant-power NEP}
\label{app:nep_time_model}

This appendix summarizes the one-dimensional constant-thrust mapping used
to compute the lower-bound NEP time-to-distance results. For fixed \(m_0\),
\(\mpay\), \(\alphatot\), and \(\Isp\), the sweep variable is the final dry
mass \(\mf=m_{\rm dry}\), equivalently the propellant fraction. The
electrical power is fixed by the dry-mass closure,
\begin{equation}
\Pe=\frac{\mf-\mpay}{\alphatot},
\qquad
\mpay < \mf < m_0,
\label{eq:appendix_power_closure}
\end{equation}
and the propellant mass is
\begin{equation}
\mprop=m_0-\mf .
\label{eq:appendix_prop_mass}
\end{equation}
With \(\ve=g_0\Isp\), constant thrust and mass flow are
\begin{equation}
\Thrust=\frac{2\eta\Pe}{\ve},
\qquad
\dot{m}=\frac{\Thrust}{\ve}.
\label{eq:appendix_thrust_mdot}
\end{equation}

During the burn,
\begin{equation}
m(t)=m_0-\dot{m}t,
\qquad
v(t)=v_0+\ve
\ln\!\left(\frac{m_0}{m(t)}\right),
\label{eq:appendix_vt}
\end{equation}
and the distance traveled is
\begin{equation}
x(t)=v_0t+
\frac{\ve}{\dot{m}}
\left[
m_0-m(t)-m(t)
\ln\!\left(\frac{m_0}{m(t)}\right)
\right].
\label{eq:appendix_xt}
\end{equation}
At burnout,
\begin{equation}
t_{\rm b}=\frac{m_0-\mf}{\dot{m}},
\qquad
v_{\rm f}=v_0+\ve\ln\!\left(\frac{m_0}{\mf}\right),
\qquad
x_{\rm b}=x(t_{\rm b}).
\label{eq:appendix_burnout}
\end{equation}
The reported arrival time to modeled distance \(D\) is therefore
\begin{equation}
\tarr(D)=
\begin{cases}
t_\star, & x(t_\star)=D,\quad 0<t_\star\le t_{\rm b},\\[4pt]
t_{\rm b}+\dfrac{D-x_{\rm b}}{v_{\rm f}}, & x_{\rm b}<D,
\end{cases}
\label{eq:appendix_arrival}
\end{equation}
where \(t_\star\) is found by a one-dimensional root solve. The tables and contours in the main text minimize \(\tarr\) over \(\mf\) subject to Eq.~(\ref{eq:appendix_power_closure}). This formulation is a rectilinear, post-injection lower-bound model; it does not include solar gravity, steering losses, finite launch energy, or the inner-solar-system escape/spiral phase.

%\bibliography{sgl_propulsion}


\end{document}